\documentclass[aps,superscriptaddress,amsmath,singlecolumn,nofootinbib,floatfix,nobalancelastpage]{revtex4-1}
\usepackage{hyperref}
\usepackage[all]{hypcap}
\usepackage{amsmath}
\usepackage{amsfonts}
\usepackage{bm}
\usepackage{graphicx}
\usepackage{grffile} % to handle dots in graphics filnames
\usepackage{color}
\usepackage{esint}
\usepackage{textpos}
\usepackage[usenames,dvipsnames]{xcolor}
\usepackage[mathscr]{eucal}
\usepackage{bm}        % for math
\usepackage{amsmath}
\usepackage{ifpdf}
\ifpdf
\usepackage[whole]{bxcjkjatype} 
\input glyphtounicode
\fi

\newcommand{\pa}{\partial}

\newcommand{\ket}{\rangle }
\newcommand{\bra}{\langle }

\newcommand{\ve}{\varepsilon}
\newcommand{\dket}{\ket\ket}
\newcommand{\dbra}{\bra\bra}
\newcommand{\up}{\uparrow}
\newcommand{\dw}{\downarrow}

\newcommand{\Vect}[1]{\mbox{\boldmath$#1$}}

\def\widebar{\accentset{{\cc@style\underline{\mskip10mu}}}}

\newcommand{\Fig}[1]{\textbf{Figure #1}}
\newcommand{\Eq}[1]{Equation #1}
\newcommand{\Sec}[1]{Section #1}

\begin{document}

% Page header
%\markboth{Oka and  Kitamura}{Floquet Engineering of Quantum Materials}

% Title
\title{
%Floquet Engineering in Ultrafast, Non-linear Electronics
Floquet Engineering of Quantum Materials
%``Floquetronics": Floquet engineered ultrafast, non-linear electronics
}

%Authors, affiliations address.
\author{Takashi~Oka}
\affiliation{$^1$ Max-Planck-Institut f{\"u}r Physik komplexer Systeme, N{\"o}thnitzer Stra{\ss}e 38, 01187 Dresden, Germany}
\affiliation{$^2$ Max-Planck-Institut f{\"u}r Chemische Physik fester Stoffe, N{\"o}thnitzer Stra{\ss}e 40, 01187 Dresden, Germany}

\author{
 Sota~Kitamura}
\affiliation{$^1$ Max-Planck-Institut f{\"u}r Physik komplexer Systeme, N{\"o}thnitzer Stra{\ss}e 38, 01187 Dresden, Germany}

%Abstract
\begin{abstract}
%Abstract text, approximately 150 words. 
\if0
Time periodic forcing in the form of coherent radiation is a standard tool for the coher- ent manipulation of small quantum systems like single atoms. In the last years, periodic driving has more and more also been considered as a means for the coherent control of many-body systems. In particular, experiments with ultracold quantum gases in optical lattices subjected to periodic driving in the lower kilohertz regime have attracted a lot of attention. Milestones include the observation of dynamic localization, the dynamic con- trol of the quantum phase transition between a bosonic superfluid and a Mott insulator, as well as the dynamic creation of strong artificial magnetic fields and topological band structures. This article reviews these recent experiments and their theoretical descrip- tion. Moreover, fundamental properties of periodically driven many-body systems are discussed within the framework of Floquet theory, including heating, relaxation dynam- ics, anomalous topological edge states, and the response to slow parameter variations.
\fi

Floquet engineering, the control of quantum systems using periodic driving, 
is an old concept in condensed matter physics, dating back to ideas such as the 
inverse Faraday effect. 
There is a renewed interest in this concept owing to the rapid developments in laser and 
ultrafast spectroscopy techniques and discovery and understanding of various ``quantum materials" 
hosting interesting exotic quantum properties. 
Here, starting from a nontechnical introduction with emphasis on the 
Floquet picture and effective Hamiltonians, 
we review the recent applications of Floquet engineering in 
ultrafast, nonlinear phenomena in the solid state. 
In particular, Floquet topological states, application to ultrafast spintronics, 
and to strongly correlated electron systems are overviewed.

\end{abstract}

%Keywords, etc.
%\begin{keywords}
%Floquet engineering, Floquet topological systems, ultrafast spintronics, 
%Mott insulator, nonequilibrium quantum systems
%\end{keywords}
\maketitle

%Table of Contents
%\tableofcontents

%%%%%%%%%%%%%%%%%%%%%%%%%%%%%%%%%%%%%%%
\section{Introduction}
%%%%%%%%%%%%%%%%%%%%%%%%%%%%%%%%%%%%%%%

How fast and drastic can we change electronic properties of materials, and what would be the most efficient way to do this? 
This is an interesting fundamental question and at the same time has connections to the 
electronic technology that supports our everyday life. 
Semiconductor devices surrounding us such as solar cells, transistors, and memories
typically involve non-equilibrium processes triggered through light-matter coupling, 
and the application of electromagnetic fields changes their properties (transport, carrier density, magnetization, etc.) drastically.
There is a growing interest in ultrafast \cite{Nasubook,Iwai06,Koshihara06,Yonemitsu08,okaLMP,Kirilyuk10,Basov11,Orenstein2012,Aoki14,Giannetti16,Mankowsky16,Basov17,Cavalleri18review} 
and non-linear \cite{Sawa08,Cario10,Pan14} electronics, aiming to find 
a way to control post-semiconductor materials hosting exotic quantum properties. 
%%%%%%%%%%%%%%%%%%%%%%%%%%%%%%%%%%%%%%%%%%%%%
\begin{figure}[t]
\includegraphics[width=13cm]{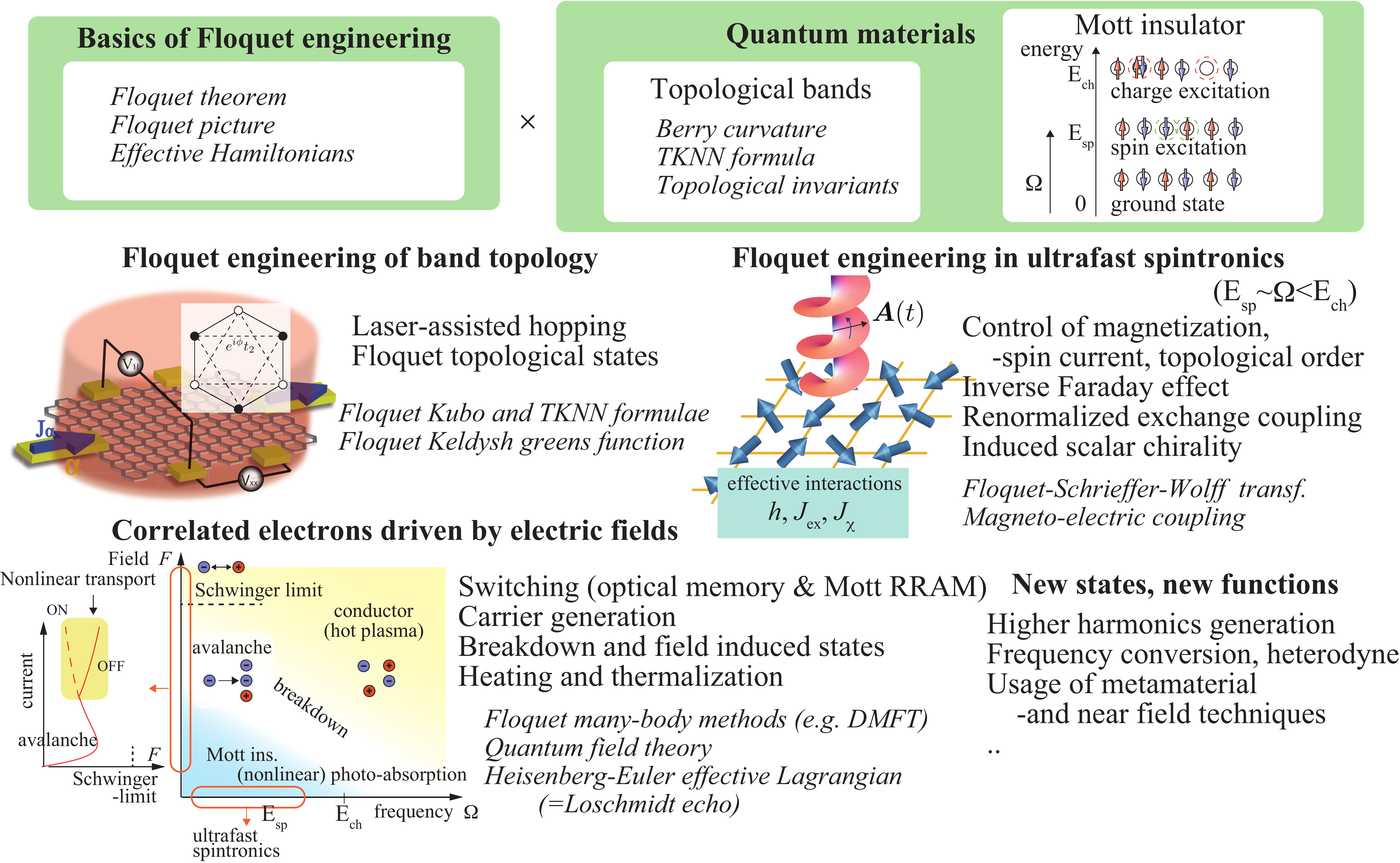}
\caption{Floquet engineering in quantum materials:  
Various processes take place when an intense laser or electric field is applied to quantum materials with 
exotic properties such as topological bands~\cite{Hasan10,RevModPhys.83.1057}, 
Dirac and Weyl semimetals \cite{NetoRevModPhys,RevModPhys.90.015001,doi:10.1146/annurev-conmatphys-031016-025458} and
strong correlation~\cite{ImadaRMP98}.
The italic keywords denote theoretical tools useful to describe them. 
}
\label{fig:overview}
\end{figure}
%%%%%%%%%%%%%%%%%%%%%%%%%%%%%%%%%%%%%%%%%%%%%

Although this research area has a long history, recently there has been some theoretical 
developments enabling us to understand various 
phenomena systematically. One powerful tool is the concept of {\it Floquet engineering}, i.e., the control 
of quantum systems using time-periodic external fields. 
Theoretically, continuous irradiation of a laser can be modeled by a time periodic perturbation
and the Hamiltonian $H(t)$ describing the irradiated system inherits the time periodicity
\begin{equation}
H(t+T)=H(t),
\label{eq:periodicity}
\end{equation}
where the periodicity $T=2\pi/\Omega$ is related to the photon energy or driving frequency $\Omega$ (we set $\hbar=1$). 
During the past decades, with the help of the Floquet theorem~\cite{Shirley65,Dunlap1986,Sambe,HanggiREVIEW1998}, which is  a temporal analogue of the Bloch theorem,
the understanding of periodically driven systems has advanced considerably especially for open transport problems \cite{Moskalets02,GRIFONI1998229}, laser driven atoms~\cite{CHU20041},
strongly correlated electron systems~\cite{Tsuji2008,Oka08Luttinger,Tsuji09,Meisner10,
Aoki14,Lee14,Mikami16,Qin17,Murakami172,Murakami17}, 
electron-phonon systems~\cite{PhysRevX.5.041050,Murakami172,Dehghani14,Babadi17}
and their universal mathematical structures for closed systems~\cite{Alessio2014,Lazarides2014,DALESSIO201319,Kuwahara16, Mori2016,Abanin,PONTE2015196,WEINBERG20171}. 
It is possible to dynamically induce 
interesting exotic quantum states
by carefully selecting the driving laser
that matches the target material. 
While Floquet engineering is now applied in several fields of physics, most notably in cold atoms in optical traps \cite{Goldman14,EckardtRMP}, here 
we focus on its application in electronic systems that is illustrated in \Fig{\ref{fig:overview}}.

%%%%%%%%%%%%%%%%%%%%%%%%%%%%%%%%%%%%%%%
\section{Basics of Floquet engineering}
\label{sec:basics}
%%%%%%%%%%%%%%%%%%%%%%%%%%%%%%%%%%%%%%%

%%%%%%%%%%%%%%%%%%%%%%%%%%%%%%%%%%%%%%%%%%%%%
\begin{figure}[tbh]
\centering 
\includegraphics[width=11.cm]{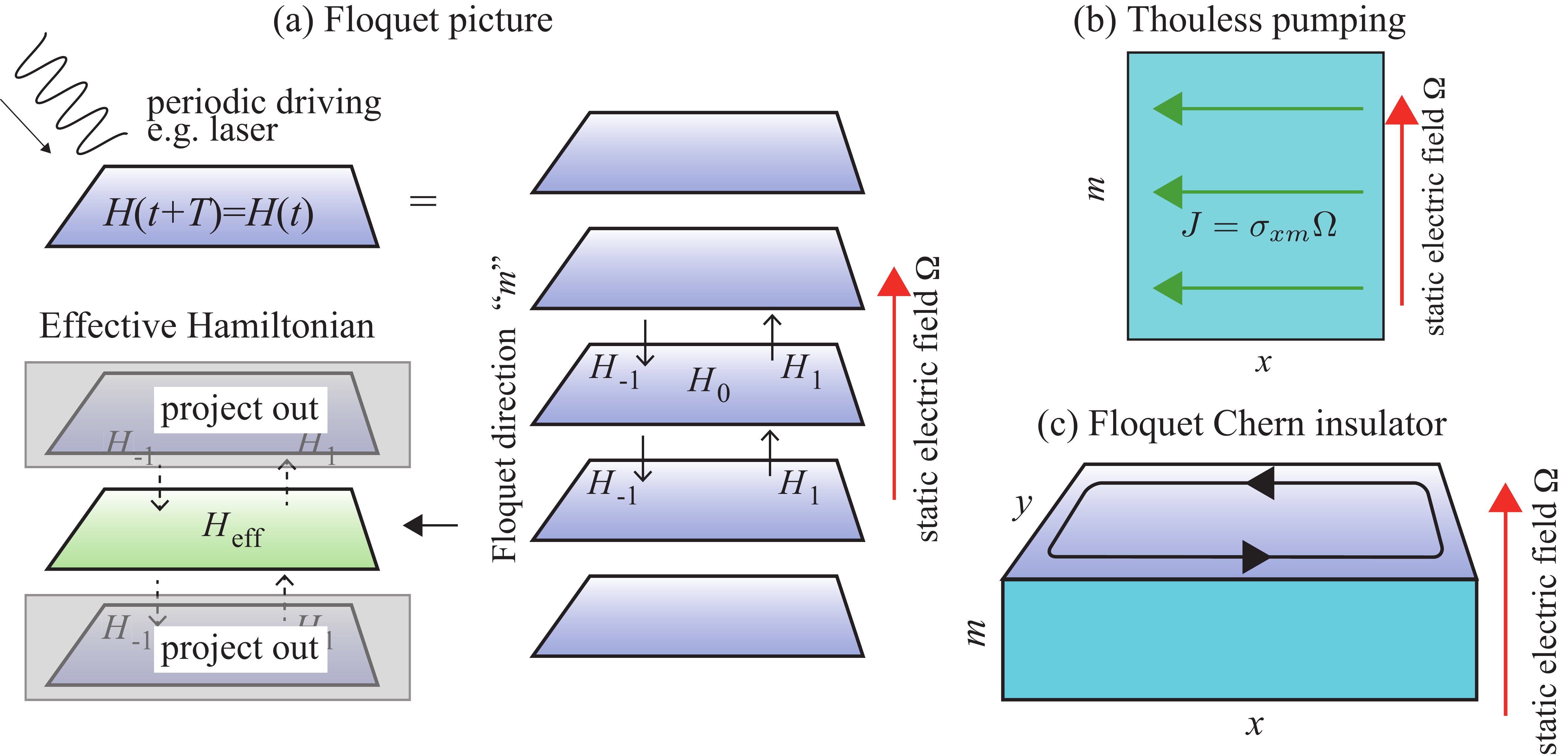}
\caption{Floquet picture for one body physics: A mapping (a) to a static higher dimensional model
gives an intuitive understanding of (b) Thouless pumping (\Sec{\ref{sec:thouless}}) and (c) Floquet Chern insulators (\Sec{\ref{sec:transport}}). The effective Hamiltonian describes physics projected on to the original Hilbert space (\Sec{\ref{sec:effectiveHamiltonians}}). }
\label{fig:floquet}
\end{figure}
%%%%%%%%%%%%%%%%%%%%%%%%%%%%%%%%%%%%%%%%%%%%
%Floquet topological phase transition takes place when a 
%quantum system is driven into a topologically nontrivial state
%by time periodic external forces. 
%Photovoltaic (photo-induced) quantum Hall effect is an 
%example of such transition and takes place when a multi-band system is 
%subject to circularly polarized laser \cite{OkaPHE09,Kitagawa2011Floquet,Lindner2011}. The external forces are not limited to  
%lasers but one can realized a topological transition by 
%many external driving. For example, in Ref.~\cite{kitagawa2}, 
%modulation of the lattice hopping parameter, which can be realized in 
%cold atoms in optical lattices, were proposed to lead to a quantum Hall state as well. In order to systematically understand the 
%nature of the phase transition, we introduce the Floquet picture of
%driven quantum states. 

We consider systems periodically driven by external fields
with a Hamiltonian satisfying the periodicity condition \Eq{\ref{eq:periodicity}}. 
%\begin{eqnarray}
%H(t+T)=H(t),
%\end{eqnarray}
%where $T$ is the time period. 
%We can find examples of time periodic systems in 
%many areas in physics. 
%In addition to laser induced dynamics, 
%Floquet theory can be applied to Thouless's adiabatic quantum pump~\cite{Thoulesspumping}, 
%which played an important role in the theory of quantum Hall effect. 
The basic idea of the Floquet method is to expand quantities into Fourier modes
$e^{-im\Omega t}$ with $m=0,\pm 1,\ldots $. 
The Floquet picture (or Shirley picture) is not only useful in doing systematic calculations, 
but also gives an intuitive way of understanding driven systems~\cite{Shirley65}.
In one-body problems, it gives a {\it mapping to a 
quantum model with one extra dimension}. 
The index $m$ of the Fourier mode %$e^{im\Omega t}$
can be considered as a lattice site index in a fictitious ``Floquet direction" (\Fig{\ref{fig:floquet}}). 
Let us see this in more details. 

%The Floquet theorem is the temporal version of the Bloch theorem:
When the Hamiltonian is time periodic, as in the Bloch theorem, 
one can take the set of solutions of the time dependent Schr\"odinger equation to be a product of a time periodic
function and a non-periodic phase factor, 
i.e., 
\begin{eqnarray}
|\Psi(t)\ket=e^{-i\ve t}|\Phi(t)\ket,\quad
|\Phi(t+T)\ket=|\Phi(t)\ket.
\label{eq:Floquetdef}
\end{eqnarray}
The time periodic function $|\Phi(t)\ket$ is called the 
Floquet state and $\ve$ the Floquet quasi-energy. 
Note that $\ve$ has an indefiniteness of integer multiples of $\Omega$.
We use the Fourier expansion
\begin{eqnarray}
H(t)=\sum_me^{-im\Omega t}H_m,\quad
|\Phi(t)\ket=\sum_me^{-im\Omega t}|\Phi^m\ket
\end{eqnarray}
for the Hamiltonian and the Floquet state. 
With this representation, the time dependent Sch\"odinger equation is mapped to an eigenvalue problem~\cite{Shirley65,Sambe}
\begin{eqnarray}
\sum_m\left(H_{n-m}-m\Omega\delta_{mn}\right)|\Phi_\alpha^m\ket=
\ve_\alpha|\Phi_\alpha^n\ket
\label{eq:Floqueteq}
\end{eqnarray}
in an extended Hilbert space.
The index $\alpha$ labels eigenstates
and $m,\;n$ are the Fourier mode indices. 
Now, the Hilbert space has been infinitely expanded, but this is compensated by the indefiniteness of $\ve$.

One can view the index $m$ as a position in the Floquet direction:
%In fact, as shown in Fig.\ref{fig:floquet}~(a), 
The system described by \Eq{\ref{eq:Floqueteq}}
is equivalent to a time independent layered one body system 
where $m$ labels the layers (\Fig{\ref{fig:floquet}}). 
%We call the extra dimension, the Floquet direction. 
The intra layer hopping is described by $H_0$,
while $H_m\;(m\ne 0)$ give inter layer couplings.  
In addition, there is a static electric field in the 
Floquet direction coming from the $m\Omega$-term
in \Eq{\ref{eq:Floqueteq}}. This fictitious electric field $\Omega$ 
plays an important role in understanding the physics of driven systems. 
For small $\Omega$, we have a lattice problem in higher 
dimensions in a weak electric field. 
If this model is a two dimensional 
Chern insulator with a Hall coefficient $\sigma_{xm}$,  
a dissipationless current $j_x=\sigma_{xm}\Omega$ 
is generated, which is nothing but the Thouless pumping~\cite{Thoulesspumping}. 
For larger $\Omega$, the layers become isolated energetically and the 
state exhibits the Wannier-Stark localization (along the Floquet direction). 
In such a situation, the high-frequency expansion is a powerful tool in 
understanding the physics systematically. %%%

%For example, in a Floquet Chern insulator, 
%circularly polarized light 
%changes the topology of a 2D trivial state 
%into a layered 2D quantum Hall state~\cite{OkaPHE09,Kitagawa2011Floquet},%
%which we will discuss in details later. 
%Classification theory of Floquet topological states %
%was proposed in Ref.~\cite{kitagawa2}, which can be 
%considered as a extension of the classification scheme of the 
%static topological states~\cite{schnyder,kitaev}. 

%Before closing the section, let us comment
%on the relation between the Floquet theory and the 
%rotating wave approximation commonly used in the field of 
%quantum optics. 
%We can obtain the rotating wave approxmation from the Floquet 
%theory by truncating the lattice in the Floquet direction, i.e,. 
%we project the theory to two sites $m=0$ and $m=1$. 
%In practice, there are many situations where we must rely on
%numerics to calculate the Floquet wave functions. In 
%such cases, we truncate the lattice in the Floquet direction
%with a cutoff, i.e., $m=-M,-M+1,\ldots,M-1,M$. We make sure that
%the result is converged as $M$ is changed. 

%%%%%%%%%%%%%%%%%%%%%%%%%%%%%%%%%%%%%%%
\subsection{Thouless pumping in the Floquet picture}
\label{sec:thouless}
%%%%%%%%%%%%%%%%%%%%%%%%%%%%%%%%%%%%%%%

%%%%%%%%%%%%%%%%%%%%%%%%%%%%%%%%%%%%%%%%%%%%%
\begin{figure}[tbh]
\centering 
\includegraphics[width=10cm]{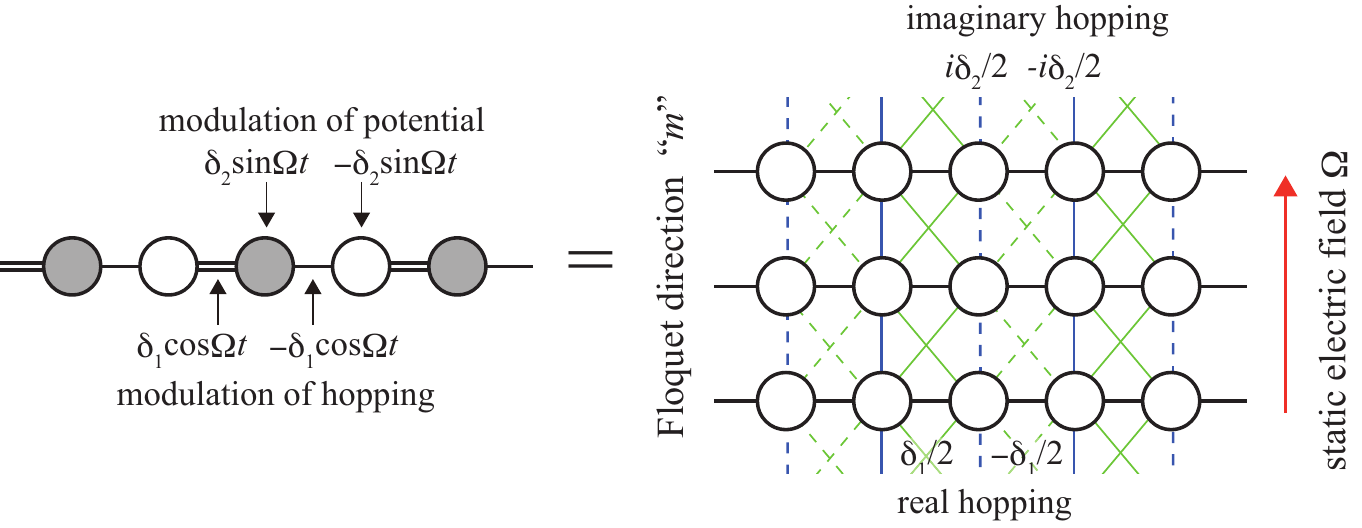}
\caption{Floquet picture for Thouless pumping in the Rice-Mele model. }
\label{fig:RiceMele}
\end{figure}
%%%%%%%%%%%%%%%%%%%%%%%%%%%%%%%%%%%%%%%%%%%%
The Thouless pumping~\cite{Thoulesspumping} is probably one of the most
well known phenomenon in time-periodic systems. 
The Rice-Mele model~\cite{RiceMele} defined by 
\begin{equation}
H(t)=-\sum_j (J+\delta_1\cos\Omega t(-1)^j)(c_{j+1}^\dagger c_j+\mbox{h.c.})+\delta_2\sin\Omega t\sum_j(-1)^jc^\dagger_jc_j
\end{equation}
is a minimum model that shows charge pumping ($J$: static hopping, $\delta_{1,2}$: modulation parameter 
of the hopping and on-site potential).  
By interpreting $\Omega t$ as momentum in the Floquet direction,
this model can be mapped to a two dimensional lattice system (\Fig{\ref{fig:RiceMele}})
governed by the eigenvalue problem \Eq{\ref{eq:Floqueteq}}, 
with $H_{0}=-J\sum_{j}(c_{j+1}^{\dagger}c_{j}+\mbox{h.c.}),\ H_{\pm1}=(1/2)\sum_{j}(-1)^{j}[-\delta_1(c_{j+1}^{\dagger}c_{j}+\mbox{h.c.})\pm i\delta_2 c_{j}^{\dagger}c_{j}]$.
With this description, the Thouless pumping 
is nothing but the two dimensional quantum Hall effect
in the $m$-$x$ plane~\cite{Kitagawa10}. 
%The induced current can be expressed by
%$J=\sigma_{xm}\Omega,$
%where frequency $\Omega=2\pi/T$ 
%emerges as the effective static electric field in the
%Floquet direction. 
The Hall conductivity of this 
effective $m$-$x$ plane is given by 
\begin{eqnarray}
\sigma_{xm}=i\sum_\lambda f_\lambda
\int_0^T\frac{dt}{2\pi}\int \frac{dk}{2\pi}
\left[
\bra\frac{\pa\psi_{\lambda }}{\pa t}|
\frac{\pa\psi_{\lambda }}{\pa k}\ket-
\bra\frac{\pa\psi_{\lambda }}{\pa k}|\frac{\pa\psi_{\lambda }}{\pa t}\ket\right],
\label{eq:sigmaxm}
\end{eqnarray}
which is nothing but the original 
expression derived by Thouless~\cite{Thoulesspumping}
and is equivalent to the TKNN (Thouless-Kohmoto-Nitingale-den Nijis) formula \cite{TKNN}
with the $y$ coordinate replaced by 
$m$.  It becomes quantized 
$(\sigma_{xm}=\frac{2\pi}{\hbar} m,\;m=0,\pm 1,\pm 2,\ldots)$
as long as the gap do not close and the distribution $f_\lambda$ is a constant 
within the band, {\em i.e.} the field strength $\Omega$ is smaller than the 
Landau-Zener tunneling threshold. 
%$\Omega$ is small enough. 
In such situation,  the relation $J=\sigma_{xm}\Omega$ states that 
integer number of charge is transferred per unit time $T=2\pi/\Omega$.

%%%%%%%%%%%%%%%%%%%%%%%%%%%%%%%%%%%%%%%
\subsection{High-frequency expansion}
\label{sec:high-frequency}
%%%%%%%%%%%%%%%%%%%%%%%%%%%%%%%%%%%%%%%
Let us consider a situation where the driving frequency $\Omega$ is 
much larger than other typical energy scales in the Hamiltonian,
so that the layers are isolated energetically.
In this situation the eigenvalue problem of \Eq{\ref{eq:Floqueteq}} can be solved efficiently 
by performing (van Vleck's) degenerate perturbation theory 
starting from the unperturbed Hamiltonian with only the $m\Omega$-term~\cite{Bukov2015,Eckardt2015}.
After performing the perturbative expansion, the eigenvalue problem becomes 
\textit{decoupled} in the Floquet direction. This means that the quasi-energy can be obtained
as eigenvalues of the static Hamiltonian.
This effective Hamiltonian has a universal form for time-periodic Hamiltonians (including many-body systems),
\begin{equation}
H_{\rm eff}^{\rm vV}=H_{0}
+\sum_{m\neq0}\left(\frac{[H_{-m},H_{m}]}{2m\Omega}+\frac{[[H_{-m},H_{0}],H_{m}]}{2m^{2}\Omega^{2}}
+\sum_{n\neq0,m}\frac{[[H_{-m},H_{m-n}],H_{n}]}{3mn\Omega^{2}}\right)
 +\mathcal{O}(\Omega^{-3}). 
\label{eq:vV}
\end{equation}
%\subsubsection{Effective Hamiltonian for free fermions}1
For later convenience, we provide the effective Hamiltonian for free fermions described by 
$H(t)=\sum_{ij}J_{ij}(t)c_{i}^{\dagger}c_{j}$
with $J_{ij}(t+T)=J_{ij}(t)=\sum_m J_{ij}^m e^{-im\Omega t}$. 
The effective Hamiltonian is calculated to be~\cite{Mikami16}
\begin{equation}
H_{\rm eff}^\text{vV}=\sum_{ij} \left( J^{(0)}_{ij} + J^{(1)}_{ij} + J^{(2)}_{ij} \right) c^\dagger_i c_j
+ \mathcal{O}( \Omega^{-3}),
\label{eq:Heff-EffectiveTightBinding-full}
\end{equation}
where the effective hopping is given as
\begin{gather}
J_{ij}^{(0)}=J_{ij}^{0}, \;
J_{ij}^{(1)}=\sum_{m\neq0}\sum_{k}\frac{J_{ik}^{-m}J_{kj}^{m}}{m\Omega},\;\nonumber\\
J_{ij}^{(2)}=\sum_{m\neq0}\sum_{kl}\left(\sum_{n\neq0}\frac{J_{ik}^{-m}J_{kl}^{m-n}J_{lj}^{n}}{mn\Omega^{2}}-\frac{J_{ik}^{0}J_{kl}^{-m}J_{lj}^{m}+J_{ik}^{-m}J_{kl}^{m}J_{lj}^{0}}{2m^{2}\Omega^{2}}\right).
\label{eq:Heff-EffectiveTightBinding}
\end{gather}

%%%%%%%%%%%%%%%%%%%%%%%%%%%%%%
\subsection{Application to many-body systems}
\label{sec:statistic}
%%%%%%%%%%%%%%%%%%%%%%%%%%%%%%
When we consider many-body systems driven periodically, we must be careful 
since the system may heat up. 
In closed periodically driven systems, a state will evolve as 
$|\psi(t)\rangle=\sum_{\alpha}c_\alpha e^{-i\ve_\alpha t}|\Phi_\alpha(t)\rangle $,
where $c_\alpha=\langle\Phi_\alpha(t_0)|\psi_0\rangle$ and $|\psi_0\ket$ is the initial state at time $t_0=0$.
$|\Phi_\alpha(t)\rangle$ are the Floquet states with quasi-energy $\ve_{\alpha}$. 
On stroboscopic time steps ($t=mT$), 
the solution $|\psi(t)\rangle$ has an identical form as the time evolution 
in static systems with the quasi-energy $\ve_\alpha$ playing the role of the (usual) energy and $|\Phi_\alpha(t_0)\rangle$ usual eigenstate.

In static nonintegrable many-body systems, for typical initial states, {\bf thermalization} is expected to occur after a sufficiently-long time evolution~\cite{Deutsch1991,Reimann2012}. 
This implies that the periodically-driven systems are also expected to thermalize because
$\bra \psi(t)|\hat{O}|\psi (t)\ket|_{t\rightarrow\infty}=\sum_\alpha e^{-\beta \ve_\alpha}\bra \Phi_\alpha (t_0)|\hat{O}|\Phi_\alpha (t_0)\ket$ should hold for an observable $\hat{O}$ in the stroboscopic time evolution. 
However, remember that $\ve_\alpha$ was only defined modulo $\Omega$, and such indefiniteness  should not appear in 
physical quantities. 
The solution to this paradox was given in References~\cite{Alessio2014,Lazarides2014}
showing that $\bra \Phi_\alpha (t_0)|\hat{O}|\Phi_\alpha (t_0)\ket$
is independent of $\alpha$. In other words,  all many-body Floquet states that 
we obtain by solving \Eq{\ref{eq:Floqueteq}} are featureless. 
This can be stated more intuitively that in closed systems, periodically driven many-body states
will thermalize to an infinite temperature state, which is consistent with a natural consequence in thermodynamics.

Nevertheless, Floquet theory is still a useful framework in many-body systems. 
{\bf Heating} should occur in a very long time scale in an appropriately chosen driving,
so that there can be a nontrivial metastable state in the shorter time scale.
%We can make the Floquet formalism to be a framework for describing this metastable state as Floquet states,
By using an appropriate perturbative expansion and truncating out 
the slow heating processes coming from higher order terms, 
we can describe this metastable state as an eigenstate of the effective Hamiltonian~\cite{Kuwahara16,Mori2016,Abanin}. 
%Such states can be described by a quench dynamics generated by an 
%effective Hamiltonian obtained by an appropriate perturbative expansion 
%truncated at a certain order~\cite{Kuwahara16, Mori2016,Abanin}. 
%Higher order terms represents perturbations and the metastable state becomes the (persistent) Floquet states 
%when the expansion is truncated at an appropriate order~\cite{Kuwahara16, Mori2016,Abanin}.
%In this situation, a periodically driven system will follow the standard 
%quench dynamics \cite{mitra2018quantum}, namely, initially there will be a coherent quantum oscillation 
%that dephases quickly reaching a Floquet prethermalized state \cite{Berges04}. 
In this situation, a periodically driven system will first ``equilibrate" with the truncated Hamiltonian,
and eventually thermalize to the true long time limit, i.e., the infinite temperature state.
The first equilibration is termed as the {\bf Floquet prethermalization},
as a special case of {\bf prethermalization}~\cite{Berges04} known 
in quench dynamics~\cite{Calabrese06,PhysRevLett.100.175702} [reviewed in \cite{Mori17,mitra2018quantum}].

%Namely, the periodically-driven state should first relax to the equilibrium state characterized by the truncated Hamiltonian (Floquet prethermalization) in a shorter time scale, and then gradually evolve to the true final state, i.e., the infinite temperature state.

%Another possibility to have nontrivial Floquet states is when the system is subject to strong disorder and is in the 
%many-body localized state. Thermalization is prevented in this case and non-trivial 
%driven states such as the discrete time crystal can be stabilized~\cite{Lazarides15}. 

Electrons in solid states are subject to various {\bf open system} relaxation processes such as
phonon scattering or coupling to an electron bath (substrate). 
When the pump is longer than their characteristic coupling time, 
the system converges to a nonequilibrium steady state, and the heating will be balanced with the relaxation \cite{Dehghani14,PhysRevX.5.041050}. 
%Also, in solid state systems with the time scale of few hundred femto seconds, 
%phonon relaxation takes place and the electron system should be considered as an
%open system.
The nontrivial steady states of such systems can be studied by the 
Floquet Greens function method combined with an appropriate many-body technique such as non-equilibrium 
dynamical mean field theory~\cite{Tsuji2008,Aoki14}. 
In \Sec{\ref{sec:correlated}}, we try to  sort out results in ultrafast and nonlinear experiments in
correlated electron systems from the point of view of many-body Floquet dynamics.

\subsection{Effective Hamiltonians}
\label{sec:effectiveHamiltonians}

While we have already introduced the concept of the effective Hamiltonian in \Sec{\ref{sec:high-frequency}},
here we discuss its general aspect in more details. %%%%%%
We have obtained the form of the effective static Hamiltonian (\Eq{\ref{eq:vV}}) by the block-diagonalization of \Eq{\ref{eq:Floqueteq}}
with regarding the inter-layer transitions as virtual processes.
If we consider this in the time domain, % rather than in the Floquet (space-time) picture,
the construction of the effective Hamiltonian can be recasted to a search for an appropriate time-periodic unitary transformation
\begin{equation}
\tilde{H}=U^\dagger(t)H(t)U(t)-iU^\dagger (t)\frac{\pa}{\pa t}U(t)
\label{eq:unitary}
\end{equation}
that makes $\tilde{H}$ time-independent.
While the existence of the transformation is assured by the Floquet theorem,
such a transformation is not unique.
One arbitrariness comes from the time-independent unitary transformation, 
by which we can generate new effective Hamiltonians once we obtain one static Hamiltonian. 
There is also an arbitrariness due to the indefiniteness of the quasi-energy.

One conventional effective Hamiltonian is the Floquet-Magnus (FM) Hamiltonian~\cite{Casas2001,Mananga2011},
which can be obtained from the time-evolution operator over a period $T$ as 
\begin{equation}
H_\text{eff}^\text{FM}=\frac{i}{T}\ln\hat{T}\exp\left[-i\int_{t_0}^{t_0+T}H(s)ds\right],
\end{equation}
where $\hat{T}$ denotes time ordering. 
While the form of the Hamiltonian changes depending on the initial time $t_0$,
those with different $t_0$ are related by unitary transformations. 
The van Vleck Hamiltonian (\Eq{\ref{eq:vV}}) is also unitary-equivalent to the FM Hamiltonian up to the truncation error.

The effective Hamiltonian depends on the perturbation schemes;
%There is a variety of choices for the unperturbed Hamiltonian.
An approximation more efficient than the high frequency expansion
can be constructed if we use the eigenbasis of $H_0-m\Omega$, the intra-layer term%
\footnote{There are also more general choices with non-block-diagonal Hamiltonians (not decoupled in the Floquet direction), which are treated in a same way after a block diagonalization of the unperturbed Hamiltonian.} 
%of \Eq{\ref{eq:Floqueteq}} 
in the Floquet picture (\Fig{\ref{fig:floquet}}),
as the starting point of perturbation. 
Note that the high-frequency expansion (\Eq{\ref{eq:vV}}) is using 
 $m\Omega$ as the unperturbed Hamiltonian.
%A more accurate expansion is obtained in a full account of $H_0-m\Omega$.
For example, if a perturbation to $H_0$ is given  by $V(t)=ve^{i\Omega t}+v^\dagger e^{-i\Omega t}$,
the second-order contribution yields
\begin{equation}
\bra a|\delta H_{\rm eff}^{\rm 2nd}|b\ket=-\sum_c\left[
\frac{\bra a|v|c\ket \bra c|v^\dagger|b\ket}{\Delta_{cb}-\Omega}+
\frac{\bra a|v^\dagger|c\ket \bra c|v|b\ket}{\Delta_{cb}+\Omega}
\right]
\label{eq:Pershan}
\end{equation}
where $H_0|a\ket=E_a|a\ket$, $\Delta _{ab}=E_a-E_b$.
This form is employed, e.g., in the theory of inverse Faraday effect~\cite{Pershan66}. %,
%and also useful for interpreting the usual Kubo formula from the viewpoint of the Floquet theory.
Application of this expansion \ref{eq:Pershan} requires all the eigenstates of $H_0$, 
which are often difficult to obtain. 
We can still use a solvable part within the Hamiltonian to construct the basis 
and perform an expansion around it. 
For example, the Floquet-Schrieffer-Wolff transformation (strong coupling expansion)
for the Hubbard model~\cite{Mentink2015,Bukov2015-2,Kitamura2017,Claassen2017}
expands in series of $1/(n_{\rm D}U-m\Omega)$ ($n_{\rm D}$ doublon number),
and the atomic Hamiltonian $U\sum_in_{i\up}n_{i\dw}-m\Omega$ is used to define the basis %as the solvable unperturbed part
(to be explained in \Sec{\ref{sec:ultrafastspintronics}}).

Among a variety of the effective Hamiltonians,
we should adopt an appropriate one in accordance with a problem of interest.
The FM Hamiltonian is directly related to the time evolution, and suitable for the initial-value problem and 
quench dynamics. 
%The $t_0$ dependence can be roughly interpreted as a dependence on the carrier-envelope phase of a recutangular pulse.
Since the FM Hamiltonian often does not respect the symmetry of the system, 
it is better to use the van Vleck expansion when we are interested in the properties of the Hamiltonian itself.
%This expansion corresponds to, in contrast to the FM Hamiltonian, 
%the situation where the external field is adiabatically inserted from the infinite past.

%The choice of the unperturbed Hamiltonian in the perturbative expansion must be 
%taken in accordance with the mechanism for the suppression of heating, 
%since the expansion becomes meaningful when the heating process is absent in low orders of the expansion.

%%%%%%%%%%%%%%%%%%%%%%%%%%%%%%%%%%%%%%%
\section{Floquet engineering of band topology}
%Transport and response under irradiation and the Floquet Chern insulator}
\label{sec:transport}
%%%%%%%%%%%%%%%%%%%%%%%%%%%%%%%%%%%%%%%

In this section, as a representative example of Floquet engineering, 
we discuss properties of the graphene irradiated by a circularly polarized laser~\cite{OkaPHE09}.
%This system exhibit a Floquet topological transition, 
%i.e., the Chern number becomes nonzero and the photo-induced Hall response is expected.
In particular, we focus on transport and response properties, and 
introduce theoretical tools to relate the nontrivial Floquet states and 
response functions.

%The two major effects of laser irradiation on electrons are
%(i) generation of photo carriers and that (ii) the electronic wave function becomes ``photo-dressed". 
%We can explore these effects systematically using the Floquet theory. 

%%%%%%%%%%%%%%%%%%%%%%%%%%%%%%%%%%%%%%%
\subsection{Laser irradiated graphene}
\label{sec:laserinducedhopping}
\label{sec:semimetal}
%%%%%%%%%%%%%%%%%%%%%%%%%%%%%%%%%%%%%%%

%%%%%%%%%%%%%%%%%%%%%%%%%%%%%%%%%%%%%%%%%%%%%
\begin{figure}[tbh]
\centering 
\includegraphics[width=10cm]{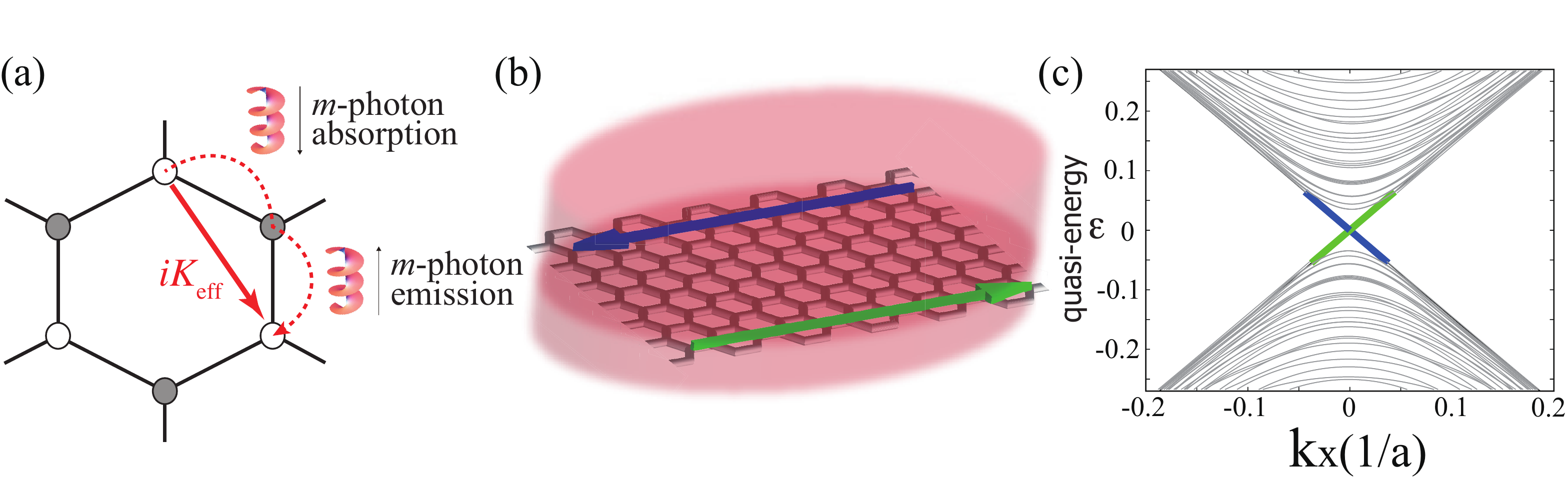}
\caption{
(a) Laser-assisted hopping  (dotted lines) for the honeycomb lattice driven by a circularly polarized laser leads to an effective 
next nearest hopping with a phase factor (\Eq\ref{eq:honeycomb-Heff}). 
Haldane's Chern insulator model \cite{Haldane88} is obtained up to first order in the high frequency expansion. 
(b) Chiral edge modes are induced under irradiation and this is seen in the (c) Floquet quasi-energy spectrum  \cite{Kitagawa2011Floquet}. 
%\textbf{\textcolor{red}{No link from text for (b,c)}}
}
\label{fig:graphene}
\end{figure}
%%%%%%%%%%%%%%%%%%%%%%%%%%%%%%%%%%%%%%%%%%%%

In graphene~\cite{NetoRevModPhys} carbon atoms form a honeycomb lattice
whose low-energy effective model is described by a tight-binding model of the $p_z$ orbital
$H = -\sum_{ij }^\text{NN} J_{ij}c^\dagger_i c_j$. Here, we take the sum over nearest-neighbor (NN) sites and $J_{ij}=J$
is the hopping amplitude. 
The energy dispersion of this model has two inequivalent Dirac cones in the Brillouin zone, 
at K and K$^\prime$ [$\bm{k}=(\pm4\pi/3\sqrt3,0)$].
The honeycomb tight-binding model in circularly polarized laser is a prototype 
of the Floquet topological insulator~\cite{OkaPHE09,Kitagawa2011Floquet,KitagawaPRB,Lindner2011}.
Let us see this by first employing the high-frequency expansion (\Sec{\ref{sec:high-frequency}}).
We introduce the laser electric field ${\bm{E}}(t)=-\partial_{t}{\bm{A}}(t)$ using the Peierls substitution as
\begin{equation}
J_{ij}(t)=
J_{ij}\exp\left(-i\int_{{\bm{R}}_{j}}^{{\bm{R}}_{i}}{\bm{A}}(t)\cdot d{\bm{r}}\right),
\label{eq:peierls}
\end{equation}
where $\bm R_i$ is the position of site $i$, and the vector potential representing a circularly polarized laser is given as 
$\bm{A}(t) = (A\cos\Omega t, A\sin\Omega t)$.
Then the Fourier components of the hopping amplitude is given as
$J^{m}_{ij}= J \mathcal{J}_{m}(A) e^{i m \phi_{ij}} $
with the bond angle $\phi_{ij}=-\tan^{-1}(R^x_i-R^x_j)/(R^y_i-R^y_j)$
and $\mathcal{J}_m$ being the $m$-th Bessel function of the first kind.
The Fourier expansion of the hopping represents laser-assisted processes, absorbing or emitting $m$-photons.
We can then use \Eq{\ref{eq:Heff-EffectiveTightBinding}} 
to obtain an effective Hamiltonian as
\begin{equation}
H_\text{eff} = -\sum_{ij}^\text{NN} J_\text{eff} c^\dagger_i c_j
+\sum_{ij}^\text{NNN}i K_\text{eff} \tau_{ij} c^\dagger_i c_j+J\mathcal{O}\left(\frac{J^2}{\Omega^2}\right),
\label{eq:honeycomb-Heff}
\end{equation}
where $\tau_{ij}$ in the next-nearest-neighbor (NNN) term takes $+1$ ($-1$) 
for the hopping with a clockwise (counterclockwise) path (from $j$ to $i$) on each hexagons.
The effective hopping amplitudes are 
\begin{equation}
J_\text{eff} = J\mathcal{J}_0(A),\;
iK_\text{eff} = -i\frac{2J^2}{\Omega}\sum_{n=1}^\infty \frac{\mathcal{J}_n^2(A)}{n}\sin\frac{2\pi n}{3}.
\label{eq:honeycomb-Keff}
\end{equation}
$J_\text{eff}$ is obtained as a time average of the original Hamiltonian, 
where a renormalization factor $\mathcal{J}_0(A)$ appears as is known as the dynamic localization~\cite{Dunlap1986,Tsuji2008}
(possibly has been observed in solids~\cite{Ishikawa14}). 
The effective NNN hopping $iK_\text{eff}$ emerges from the two-step laser-assisted hopping process \Fig{\ref{fig:graphene}(a)}.
\Eq{\ref{eq:honeycomb-Keff}} is nothing but the Hamiltonian of the Haldane model~\cite{Haldane88}, and at the Dirac cones K and K$^\prime$ 
the model exhibits a topological gap opening
leading to a nontrivial Chern number. 
Chiral edge modes emerges in the quasi-energy spectrum \Fig{\ref{fig:graphene}(b),(c)},
and their direction as well as the sign of the Chern number depends on the sign of $K_\text{eff}$
that can be flipped by changing the polarization $\Omega\to-\Omega$. 
%%%%%%%%%%%%%%%%%%%%%%%%%%%%%%%%%%%%%%%%%%%%%
\begin{figure}[tbh]
\centering 
\includegraphics[width=8.cm]{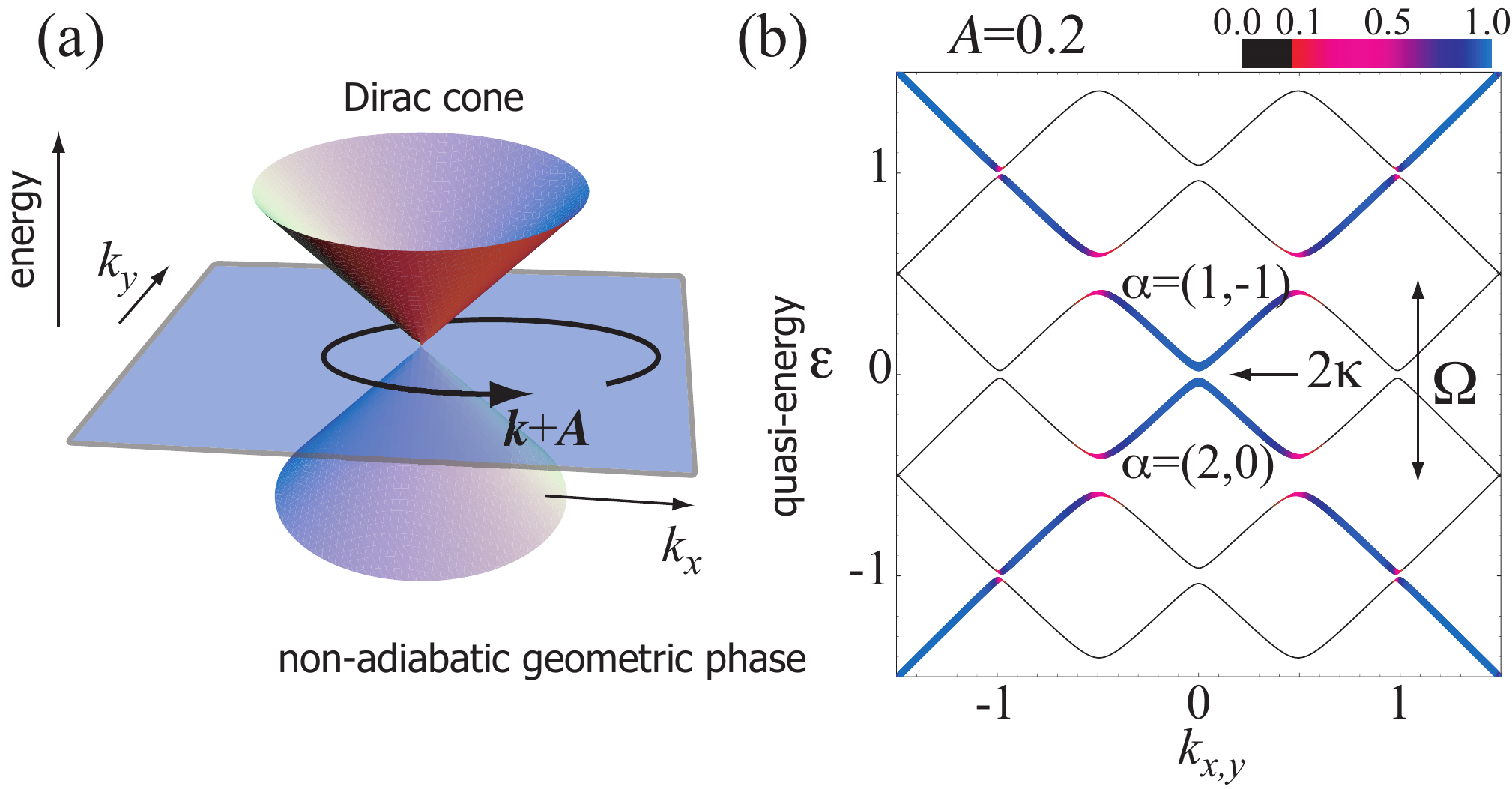}
\caption{(a) States with momentum $\Vect{k}$ acquires a geometric phase near the Dirac node. 
(b) Floquet spectrum of two dimensional Dirac system in circularly polarized laser. Color coding represents the static weight (adapted from Reference~\cite{OkaPHE09}).   }
\label{fig:Floquetspectrum}
\end{figure}
%%%%%%%%%%%%%%%%%%%%%%%%%%%%%%%%%%%%%%%%%%%%

While the high-frequency expansion gives an intuitive picture for the Floquet topological phase transition,
let us comment on the role played by the geometric phase on the gap opening. 
In momentum space, Through the Peierls substitution (\Eq{\ref{eq:peierls}}), 
a state with momentum $\bm{k}$  start to move as $\bm{k}+\bm{A}(t)$ in the momentum space.
For the circularly polarized laser, this motion is circular (\Fig{\ref{fig:Floquetspectrum}(a)}), and 
the state acquires the Aharonov-Anandan phase~\cite{AharonovAnandan} (or the non-adiabatic extension of the Berry phase~\cite{Berry45}) 
during the time evolution.
The Floquet quasi-energy is written as 
\begin{eqnarray}
\ve_\alpha=\dbra\Phi_\alpha|H(t)|\Phi_\alpha\dket
+\Omega \gamma^{\rm AA}_\alpha/2\pi,
\end{eqnarray}
a sum of the dynamical phase and the Aharonov-Anandan phase.
We have used the time averaged inner product and matrix element of a time dependent operator as
$\dbra \Phi_\alpha|\Phi_\beta\dket
\equiv \frac{1}{T}\int_0^Tdt\bra \Phi_\alpha(t)|
\Phi_\beta (t)\ket=\sum_m\bra\Phi_\alpha^m|\Phi_\beta^m\ket,$
and $
\dbra \Phi_\alpha|\mathcal{O}|\Phi_\beta\dket
=\frac{1}{T}\int_0^Tdt\bra \Phi_\alpha(t)|\mathcal{O}(t)|
\Phi_\beta (t)\ket$, respectively. 
At the Dirac points  K and K$^\prime$ [$\bm{k}=(\pm4\pi/3\sqrt3,0)$]
 (with a linearized dispersion), the Aharonov-Anandan phase
becomes~\cite{OkaPHE09}
\begin{eqnarray}
\gamma^{\rm AA}_\alpha\equiv T\dbra\Phi_\alpha|i\pa_t|\Phi_\alpha\dket=\pm \pi\left\{
[4(A/\Omega)^2+1]^{-1/2}-1
\right\}.
\end{eqnarray}
In the adiabatic limit ($\Omega\to 0$),
it converges to the Berry phase $\mp\pi$. 
The size of the gap $2\kappa$ can be evaluated as 
\begin{eqnarray}
2\kappa=\sqrt{4A^2+\Omega^2}-\Omega.
\label{eq:gap}
\end{eqnarray}
In solid states, the gap opening of Dirac node as well as the other replica bands shown in \Fig{\ref{fig:Floquetspectrum}(b)}
was observed in a time resolved ARPES experiment~\cite{Wang453}. 
In artificial matters, the periodically driven graphene 
was simulated in cold atoms in optical lattices~\cite{Jotzu2014b}
as well as in photonic wave guides~\cite{Rechtsman2013}, realizing the Haldane model (Floquet Chern insulator).  
%%%%%%%%%%%%%%%%%%%%%%%%%%%%%%%%%%%%%%%
\subsection{Floquet Kubo and TKNN formulae}
\label{sec:FloquetKuboformula}
%%%%%%%%%%%%%%%%%%%%%%%%%%%%%%%%%%%%%%%

%%%%%%%%%%%%%%%%%%%%%%%%%%%%%%%%%%%%%%%%%%%%%
\begin{figure}[tbh]
\centering 
\includegraphics[width=12.5cm]{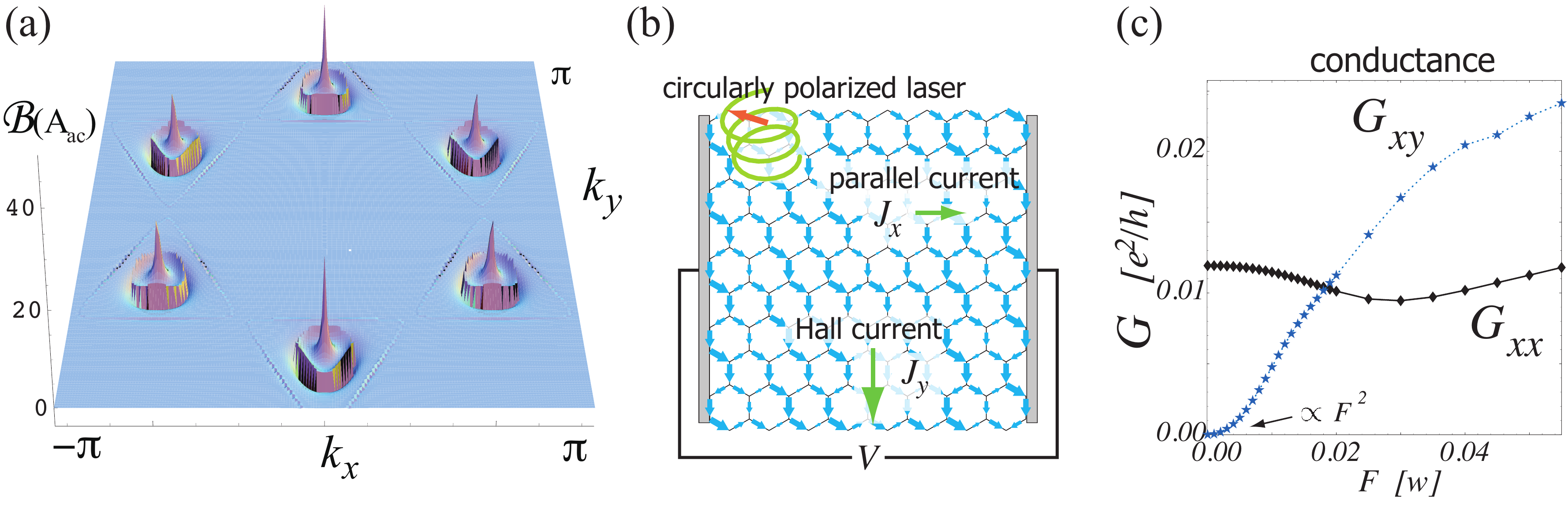}
\caption{
(a) Berry curvature of the Floquet band in graphene irradiated by circularly polarized laser (adapted from Reference~\cite{OkaPHE09}).
Detection of Floquet Chern insulator can be done with (b)  transport measurements of the
(c) Hall conductance $G_{xy}$~\cite{OkaPHE09}. ($F=\Omega A_{\rm ac}$). }
\label{fig:photoBerry}
\end{figure}
%%%%%%%%%%%%%%%%%%%%%%%%%%%%%%%%%%%%%%%%%%%%
In equilibrium, the quantum Hall effect at zero temperature is related to the Chern number through the TKNN formula. 
In a similar way, we need the Floquet-Kubo formula~\cite{Torres05} and the Floquet TKNN formula~\cite{OkaPHE09}
to characterize the response of Floquet topological states. Here, we discuss nonlinear optical and transport properties in the presence of irradiation. 
When a {\it strong} laser is applied ($A_{\rm ac}$) to a system, its
response to an additional {\it weak} perturbation can be changed. 
The response function is defined by a linear relation 
\begin{eqnarray}
j_i(\omega)=\sigma_{ij}(A_{\rm ac};\omega)E_j(\omega)
\end{eqnarray}
between the weak probe electric field $E_j$
and the induced current $j_i$. 
The effect of the external AC-field is
taken into account using Floquet states 
as the basis.
The optical response function for irradiated systems
in the DC-limit ($\omega\to 0$)  becomes
\begin{eqnarray}
\sigma_{ij}(\Vect{A}_{\rm ac})=
i\int\frac{d\Vect{k}}{(2\pi)^d}\sum_{\alpha\in BZ_1}\sum_{\beta\ne\alpha}
\frac{f_\beta(\Vect{k})-f_\alpha(\Vect{k})}
{\ve_\beta(\Vect{k})-\ve_\alpha(\Vect{k})}
\frac{\dbra\Phi_\alpha(\Vect{k})|J_j|
\Phi_\beta(\Vect{k})\dket\dbra\Phi_\beta(\Vect{k})|
J_i|\Phi_\alpha(\Vect{k})\dket}
{\ve_\beta(\Vect{k})-\ve_\alpha(\Vect{k})+i0^+},
\end{eqnarray}
which was given in References~\cite{OkaPHE09,Kitagawa2011Floquet,Dehghani15}.
Here,
$\ve_\alpha$ is the quasi-energy of Floquet states $\alpha$, and
$f_\alpha$ is its occupation fraction. 
Note that the index $\alpha$ is limited to the first Brilliouin zone ($\ve_{\alpha}\in [-\Omega/2,\Omega/2)$)
whereas $\beta$ is taken over the whole Floquet spectrum which is extended from 
the original spectrum by the inclusion of photon-dressed replica states. 
The Hall coefficient
is given by the Floquet-TKNN formula \cite{OkaPHE09}
\begin{eqnarray}
\sigma_{xy}(\Vect{A}_{\rm ac})=
e^2\int\frac{d\Vect{k}}{(2\pi)^2}
\sum_{\alpha\in BZ_1} f_\alpha(\Vect{k})\left[
\nabla_{\Vect{k}}\times\Vect{\mathcal{A}}_\alpha(
\Vect{k})
\right]_z.
\end{eqnarray}
The artificial gauge field and its associated Berry curvature is defined by 
\begin{eqnarray}
\Vect{\mathcal{A}}_\alpha(\Vect{A}_{\rm ac};
\Vect{k})=-i\dbra\Phi_\alpha(\Vect{k})|\nabla_{\Vect{k}}|
\Phi_\alpha(\Vect{k})\dket,\;
\Vect{\mathcal{B}}_\alpha(\Vect{A}_{\rm ac};\Vect{k})=
\nabla_{\Vect{k}}\times\Vect{\mathcal{A}}_\alpha(
\Vect{k}).
\end{eqnarray}
\Fig{\ref{fig:photoBerry}(a)}
shows the Berry curvature of a Floquet band, where 
we can see that peaks appear at the Dirac points. 
There are also concentrical patterns around the 
Dirac points where resonant hybridization occurs between the bands. 
There are two promising detection schemes 
of the Hall effect, which is the transport measurement in the presence of 
irradiaiton (\Fig{\ref{fig:photoBerry}(b,c)} illustrates ), and the time resolved Kerr effect~\cite{Oka2010g,Dehghani15}.  
In the Kerr effect, the polarization angle shifts as~\cite{Ikebe10}
\begin{eqnarray}
\Theta_H
\if0=\frac{1}{2}\mbox{arg}\left[
\frac{n_0+n_s+(\sigma_{xx}+i\sigma_{xy}/(c\ve_0)}
{n_0+n_s+(\sigma_{xx}-i\sigma_{xy}/(c\ve_0)}
\right]
=\frac{1}{(n_0+n_s)c\ve_0}\sigma_{xy}(\omega)
\fi
\sim (\sigma_{xy}\mbox{ in units of }\frac{e^2}{h})\;\times\; 6.3\;\mbox{mrad}
\end{eqnarray}
in the presence of a quantum Hall state. 
The Hall conductivity is not necessarily quantized in a Floquet Chern insulator 
since the electrons are photo-excited and the distribution function  $f_\alpha$
is not the simple zero-temperature Fermi distribution. 
However, there has been several numerical analysis finding conditions for
a quantized Hall coefficient in open systems~\cite{GuAuerbach11,Dehghani15}．

%%%%%%%%%%%%%%%%%%%%%%%%%%%%%%%%%%%%%%%
\section{Floquet engineering in ultrafast spintronics }
\label{sec:ultrafastspintronics}
%%%%%%%%%%%%%%%%%%%%%%%%%%%%%%%%%%%%%%%

%%%%%%%%%% Fig %%%%%%%%%%
\begin{figure}[thb]
\centering
\includegraphics[width=13cm]{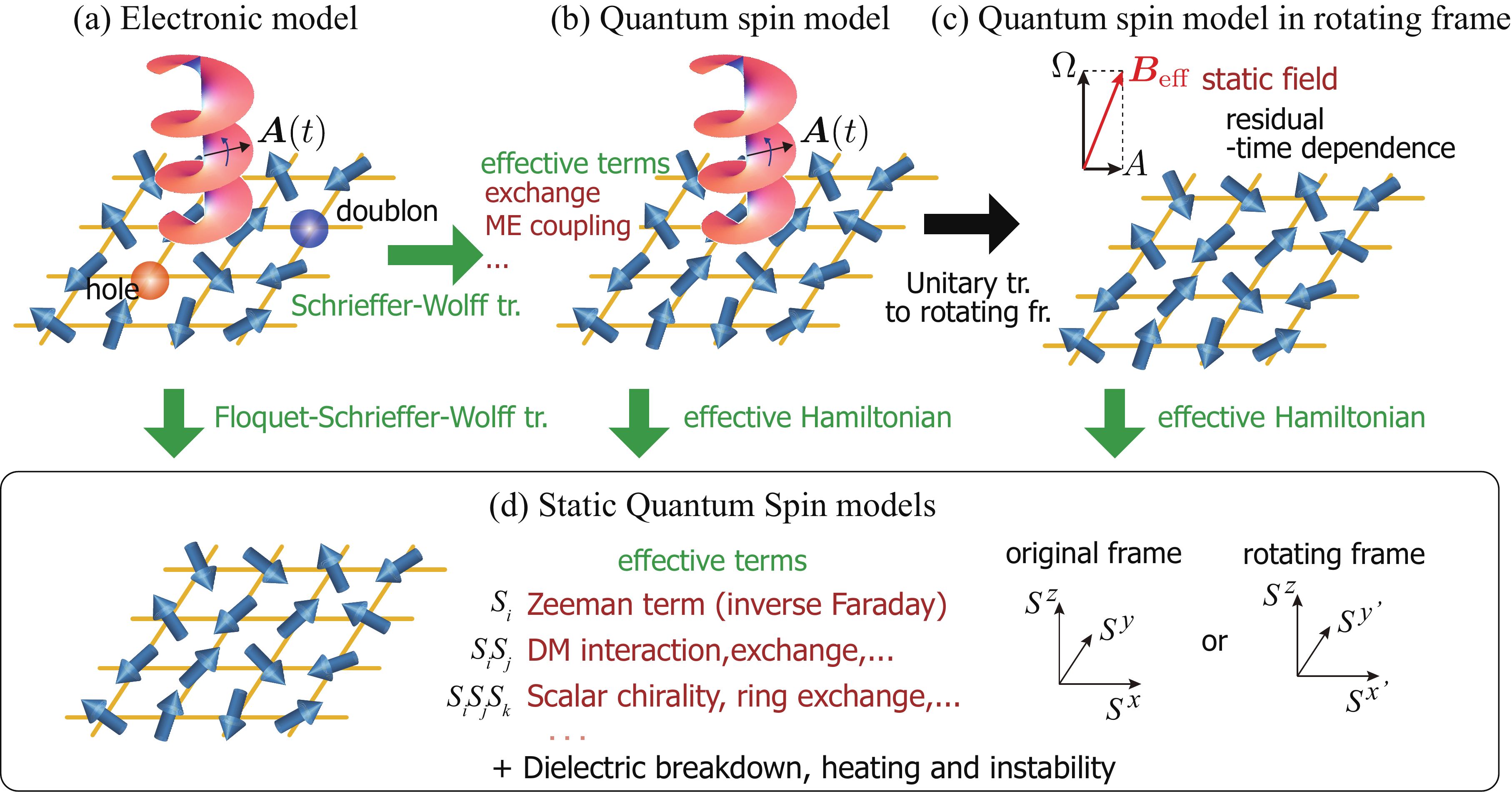}
\caption{
Effective theories for Floquet engineering in ultrafast spintronics: 
(a) The starting system is the electronic models, e.g., band ferromagnets and Mott insulators, under irradiation. 
(b) Time dependent spin model is obtained by assuming that the field is weak and $\Omega$ much smaller than the charge gap. 
(c) For circularly polarized laser, moving to the rotating frame gives a 
static magnetic field. 
(d) Effective Hamiltonians can be obtained by using appropriate expansion schemes. 
}
\label{fig:framework}
\end{figure}
%%%%%%%%%%%%%%%%%%%%
Spintronics is a new branch of electronics where the spin degrees of freedom is used to carry and store information
via spin current and magnetization~\cite{Zutic04}. 
In ultrafast spintronics, laser is used to 
control spins and magnetism in the time scale of pico-seconds or faster
\cite{Kimel05,Stanciu07,Kirilyuk10}. 
A direct way to access spins is by the Zeeman coupling that couples to the magnetic field component of laser. 
Another relevant coupling is the magneto-electric (ME) coupling which arrows 
the electric field to interact with the polarization that depends on the spins.  
Several types of ME couplings are 
proposed [reviewed in \cite{Tokura14}] such as (i) inverse Dzyaloshinskii-Moriya  (DM) model~\cite{Tanabe65,Katsura}
$\Vect{P}\propto \Vect{e}_{\Vect{r},\Vect{r}'}\times (\Vect{S}_{\Vect{r}}\times\Vect{S}_{\Vect{r'}})$
and (ii) exchange striction model
$\Vect{P}\propto \Vect{\pi}_{\Vect{r},\Vect{r}'}\times (\Vect{S}_{\Vect{r}}\cdot\Vect{S}_{\Vect{r'}})$.

The idea of Floquet engineering and effective Hamiltonians can be applied to ultrafast spintronics. 
Since the spin degrees of freedom originates from electrons, 
the construction of the effective Hamiltonian has different levels of approximation illustrated in \Fig{\ref{fig:framework}}.

\noindent
\textbf{(a) Electronic model}:\;
If direct electronic excitations are involved, we should start the construction from the electronic model. 
A classic example is the inverse Faraday effect studied by Pershan {\em et al.} \cite{Pershan66}, where 
they used \Eq{\ref{eq:Pershan}} to obtain
an effective Zeeman coupling 
%\begin{eqnarray}
$\delta H_{\rm eff}=h_{\rm eff}S_z$
%\end{eqnarray}
for electron systems in circularly polarized laser applied along the $z$-axis. 
The effective magnetic field $h_{\rm eff}$, obtained by considering virtual electronic excitations, is
a nonlinear function of the laser strength and frequency and can be extremely large~\cite{Kimel05,Stanciu07,Kirilyuk10}. 
On the other hand, in strongly correlated insulators such as the Mott insulator, 
using the second order perturbation  \Eq{\ref{eq:Pershan}} is difficult, and thus
it is plausible to combine the Floquet picture with the standard Schrieffer-Wolff 
transformation~\cite{Mentink2015,Bukov2015-2,Kitamura2017,Claassen2017}, which will be explained in Section \ref{sec:FloquetSW}. 

\noindent
\textbf{(b) Quantum spin model}:\;
Another way to construct the effective theory is to start from 
quantum spin models
including the coupling to laser fields~\cite{Sato14,Sato16}, i.e.,
\begin{equation}
 H(t)=H_{0}
   -g\mu_{\rm B}\bm{B}(t)\cdot\bm{S}-\bm{E}(t)\cdot\bm{P}. 
\label{eq:TimeDepMultiferro}
\end{equation}
Here, $H_0$ denotes the original quantum spin models, e.g., Heisenberg model, and 
the other terms are the Zeeman and ME coupling. This approach is 
justified when the field is much slower than the electron dynamics ($\Omega\ll E_{\rm ch}$). 
An effective Hamiltonian for circularly polarized laser based on the high frequency expansion (\Eq{\ref{eq:vV}})
\begin{align}
\delta{\cal H}_{\rm eff}
   &=-\frac{i}{2\Omega}\big\{
        \beta^{2}[S^{x},S^{y}]
     +\alpha\beta([\tilde P^{x},S^{x}]+[\tilde P^{y},S^{y}])      
     +\alpha^{2}[\tilde P^{x},\tilde P^{y}]
     \big\}
\label{eq:synthetic}\\
&\;\sim \mbox{1 spin term }+\mbox{2 spin term}+\mbox{3 spin term}
%&\sim \mbox{3 spin term ("chirality")}+\mbox{2 spin term ("DM")}+\mbox{1 spin term ("Zeeman")}
\end{align}
can be obtained~\cite{Sato16}. Here,  the dimensionless polarization is defined by 
$\tilde{\bm P}={\bm P}/g_{\rm me}$ with 
$g_{\rm me}$ being the ME coupling constant and 
$\alpha=g_{\rm me}E_{0}$ and $\beta=g\mu_{\rm B}E_{0}c^{-1}$. 
Evaluating the commutators, we obtain terms that are product of 1, 2, 3 spins, for example the Zeeman, DM and spin chirality,
and their precise forms depend on the ME coupling reflecting the symmetry of the crystal.

\noindent
\textbf{(c) Quantum spin model in the rotating frame}:\;
In the special case where the applied field is circularly polarized, 
we can obtain an exact form of the effective Hamiltonian without carrying out the perturbative expansion:
A quantum spin system in rotating magnetic field described by 
$H(t)=H_0+A\cos(\Omega t)S_x +A\sin(\Omega t)S_y$
can be mapped to a simple static magnetization problem $\tilde{H}=H_0+\Omega S_z+AS_x$
by moving to the rotating frame using a unitary transformation $U(t)=\exp(i\Omega S_z t)$
assuming that $H_0$ is rotationally symmetric around the $z$-axis.
The effective magnetic field $h_{\rm eff}=(A,0,\Omega)$ 
acting on the mapped system can be large, e.g., $\Omega$ in teraheltz corresponds to 
few teslas, and can be used to orient the spins and induce magnetization\cite{Takayoshi14-1,Takayoshi14-2}. 

%%%%%%%%%%%%%%%%%%%%%%%
\subsection{Floquet-Schrieffer-Wolff transformation }
\label{sec:FloquetSW}
%%%%%%%%%%%%%%%%%%%%%%%%%%%%%%%%%%%%%%%
Here, we will see how the laser electric field influences the spin degree of freedom
via the framework of the strong-coupling expansion for the periodically-driven Hubbard model.
We introduce the Hubbard model under a laser electric field,
\begin{equation}
H(t)=T(t)+UD=-\sum_{ij\sigma}t_{ij}e^{-iA_{ij}(t)}{c}_{i\sigma}^{\dagger}{c}_{j\sigma}+U\sum_{i}n_{i\uparrow}n_{i\downarrow},
\label{eq:tdHubbard}
\end{equation}
where $t_{ij}$ describes hopping of electrons $c$ from site $j$ to $i$, 
while $U$ the on-site density interaction. 
The oscillating electric field is described by the Peierls phase $A_{ij}(t)$.

First, let us quickly review the strong-coupling expansion of the Hubbard model in the static case.
In the strong coupling limit of the Hubbard model $t_{ij}=0$,
the ground states are any electron configurations with no double occupancy, 
which have a macroscopic degeneracy.
In particular, the subspace spanned by the ground states at half filling is described by spin configurations.
The macroscopic degeneracy lifts when we introduce the hopping $t_{ij}\neq0$ as a perturbation.
The low-energy effective model that describes this lift is the Heisenberg model, where
the antiferromagnetic exchange interaction $J_{ij}$ is given as $J_{ij}=4t_{ij}^2/U$.

To perform a perturbative expansion of macroscopically degenerate systems in a systematic manner,
it is convenient to employ the canonical transformation (Schrieffer-Wolff transformation).
Namely, we consider a unitary transformation $e^{iS}$ as a series in $t_{ij}$,
and determine it order by order to block-diagonalize the Hamiltonian with respect to double-occupancy $D$,
which classifies the eigenstates in the atomic limit.
We then obtain the effective spin Hamiltonian as the $D=0$ sector of the transformed Hamiltonian.
Our goal in this section is to see the influence of an external field on the Mott insulator 
described by this strong-coupling expansion.

One can extend this scheme to a time-dependent situation by considering a time-dependent transformation.
We first perform a series expansion of the transformed Hamiltonian $H^\prime(t)$ in $S(t)$ as~\cite{Kaminski00,Goldin00}
\begin{eqnarray}
H^\prime(t)&=&e^{iS(t)}H(t)e^{-iS(t)}-e^{iS(t)}i\partial_te^{-iS(t)}\nonumber\\
&=&H(t)+[iS(t),H(t)-i\partial_t]+\frac{1}{2}[iS(t),[iS(t),H(t)-i\partial_t]]+\cdots,
\end{eqnarray}
and further expand $S(t)$ in $t_{ij}$ as $S(t)=S^{(1)}(t)+S^{(2)}(t)+\cdots.$
Then the first-order term of $H^\prime(t)$ is given as
\begin{equation}
H^{\prime(1)}=T(t)+[iS^{(1)}(t),UD]-\partial_tS^{(1)}(t).
\end{equation}
To block-diagonalize $H^\prime(t)$, we determine $S$ by solving a set of differential equations
\begin{equation}
\partial_tS^{(1)}_{+d}(t)=-idUS^{(1)}_{+d}(t)+T_{+d}(t)
\end{equation}
for $d\neq0$, where $S_{+d}$, $T_{+d}$ denote terms which increase double occupancy $D$ by $d$.
While one can solve these for arbitrary fields with an appropriate boundary condition,
here we consider a monochromatic laser $A_{ij}(t)=A_{ij}\cos(\Omega t-\phi_{ij})$
in particular, and consider a time-periodic solution.
By carrying out the second-order perturbation, we obtain the Heisenberg model with a modified 
exchange interaction~\cite{Mentink2015}
\begin{equation}
J_{ij}(t)=\sum_{m,n}(-1)^{m}\frac{4|t_{ij}|^{2}\mathcal{J}_{n+m}(A_{ij})\mathcal{J}_{n-m}(A_{ij})}{U-(m+n)\Omega}\cos2m(\Omega t-\phi_{ij}),
\end{equation}
where $\mathcal{J}_m$ is $m$th Bessel function. 
This is the lowest-order contribution of the electric field on the spin interaction.
The obtained spin interaction is time periodic, and the effective static Hamiltonian 
can be obtained by performing the high-frequency expansion (\Sec{\ref{sec:high-frequency}}).

While we have considered a simple spin interaction in the Hubbard model here, 
there are various mechanism to produce effective spin interactions via virtual processes,
such as the Kugel-Khomskii coupling in the multi-band systems~\cite{Kugel1982,Eckstein2017}, 
superexchange coupling in multiferroic systems~\cite{Katsura}, and 
anisotropic spin coupling under the strong spin-orbit coupling~\cite{Chaloupka2010,Bhattacharjee2012}, 
for all of which we can apply the described scheme.

As we have mentioned in \Sec{\ref{sec:statistic}},
the heating processes in higher order perturbation are (implicitly) truncated out in the present scheme.
Such processes associated with charge excitations emerge as a divergence of the expansion 
(due to a vanishing energy denominator $DU-m\Omega$).
This divergence originally comes from an additional degeneracy between sectors with different $D$ in the atomic limit%
\footnote{For example, when $U/\Omega=p/q$ with coprime $p,q$, 
the quasienergy $\epsilon=DU-m\Omega$ is zero not only for $(D,m)=(0,0)$
but also for $(q,p),(2q,2p),\ldots$.}.
Hence, we have to be aware that, in the true degenerate perturbation theory,
one cannot block-diagonalize the Hamiltonian with $D$ (nor $m$), and the above scheme is valid up to a certain finite order~\cite{Kitamura2017}.
The inter-sector (changing $D$) terms is nothing but the charge excitations, which leads to the heating.
It is a general property of the effective Hamiltonian approach that 
the heating processes emerge as a divergence of the series expansion, 
and the error due to the truncation of the expansion gives a finite (but very long in many cases) lifetime~\cite{Kuwahara16,Mori2016,Abanin,EckardtRMP}.

%%%%%%%%%%%%%%%%%%%%%%%%%%%%%%%%%%%%%%%
\subsubsection{Control and detection of spin chirality using laser}
\label{sec:chrality}
%%%%%%%%%%%%%%%%%%%%%%%%%%%%%%%%%%%%%%%
%%%%%%%%%% Fig %%%%%%%%%%
\begin{figure}[thb]
\centering
\includegraphics[width=12cm]{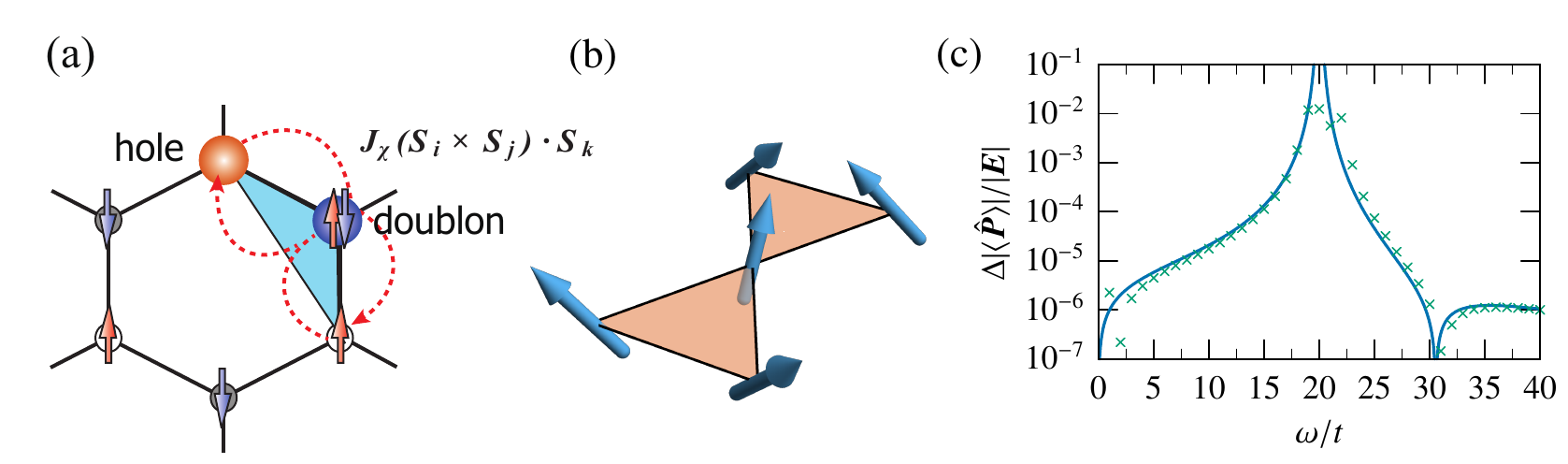}
\caption{
(a) Laser assisted virtual hopping process leading to the scalar spin chirality term.
(b) classical configuration of spins with nonzero scalar chirality, $(\hat{\bm{S}}_{i}\times\hat{\bm{S}}_{j})\cdot\hat{\bm{S}}_{k}$.  
(c) Circular dichroism: Difference of the induced electric polarization between 
left and right circularly-polarized lasers, as a function of the photon energy. 
Crosses are numerical results for a three-site Hubbard cluster, and
the solid curve is obtained from \Eq{\ref{eq:dielectric-function}}.
}
\label{fig:chirality}
\end{figure}
%%%%%%%%%%%%%%%%%%%%

While we have confirmed the modification of the exchange interaction,
more intriguing possibility is to induce an emergent interaction term that is absent 
in the static model by irradiating a laser.
This can be achieved by applying a field which breaks some symmetry of the original system.
For instance, the time-reversal symmetry is broken when the circularly-polarized laser is applied.
If we continue the strong-coupling expansion to the fourth-order \Fig{\ref{fig:chirality}(a)}, 
the emergent term is the scalar spin chirality term $\chi_{ijk}=\bm{S}_i\cdot(\bm{S}_j\times\bm{S}_k)$.
A classical spin configuration with a nonzero scalar chirality is shown in \Fig{\ref{fig:chirality}(b)} for illustration.
Upon irradiation, a effective scalar chirality term $\delta H_{\rm eff}=\sum_{ijk}J_{\chi,ijk}\chi_{ijk}$ with
\begin{align}
J_{\chi,ijk}=&-4|t_{ij}|^{2}|t_{jk}|^{2}\sum_{l,n}\sum_{m\neq0}
\biggl[\frac{\mathcal{J}_{l+m}(A_{ij})\mathcal{J}_{l}(A_{ij})\mathcal{J}_{n+m}(A_{jk})\mathcal{J}_{n}(A_{jk})\sin m(\phi_{ij}-\phi_{jk})}{(U-l\Omega)(U-n\Omega)(U-(l+n+m)\Omega)}\nonumber\\
&+\frac{\mathcal{J}_{l+m}(A_{ij})\mathcal{J}_{l-m}(A_{ij})\mathcal{J}_{n+m}(A_{jk})\mathcal{J}_{n-m}(A_{jk})\sin2m(\phi_{ij}-\phi_{jk})}{m\Omega(U-(l+m)\Omega)(U-(n+m)\Omega)}\biggr]
\label{eq:Jchi}
\end{align}
emerges\cite{Claassen2017,Kitamura2017}.
This opens an intriguing possibility of Floquet engineering  exotic quantum phases such as a 
chiral spin liquid phase\footnote{Floquet chiral spin liquid with Majorana edge modes is proposed in 
Kitaev systems~\cite{KITAEV20062} under circularly polarized laser~\cite{Sato14}. 
This theory starts from the driven quantum spin model. 
}~\cite{Bauer2014}.

The light-induced interaction has a potential application as a new probe:
For the above example, the circularly-polarized laser acts as a conjugate field to the scalar spin chirality
for general Mott insulators, as in the magnetic field conjugate to the spin.
Namely, the coupling constant for the scalar chirality is reduced to be
\begin{equation}
J_{\chi,ijk}\sim\frac{2\mathcal{A}_{ijk}|t_{ij}|^{2}|t_{jk}|^{2}\Omega(7U^{2}-3\Omega^{2})}
{U^{2}(U^2-\Omega^2)^{3}}
i(\bm{E}^{\ast}\times\bm{E})_z,
\label{eq:JchiE2}
\end{equation}
in the leading order of the field amplitude, where $\mathcal{A}_{ijk}$ is the area enclosed by sites $i,j,k$.
The interaction term describes the modulation of the dielectric function proportional to the scalar chirality,
when it is seen as a term in the Hamiltonian of the electromagnetic field.
From \Eq{\ref{eq:JchiE2}}, the modulation is obtained as 
an imaginary off diagonal part leading to the circular dichroism~\cite{Kitamura2017}
\begin{equation}
\epsilon_{xy}(\omega)=i\sum_{ijk}\frac{4|t_{ij}|^{2}|t_{jk}|^{2}\omega(7U^{2}-3\omega^{2})}{U^{2}(U^{2}-\omega^{2})^{3}}\mathcal{A}_{ijk}\left\langle (\hat{\bm{S}}_{i}\times\hat{\bm{S}}_{j})\cdot\hat{\bm{S}}_{k}\right\rangle.
\label{eq:dielectric-function}
\end{equation}
Namely, one can read out the presence of the scalar chirality via the circular dichroism.
\Fig{\ref{fig:chirality}(c)} shows the difference of the dielectric function 
between circularly polarized light with a different chirality.

\noindent
\textbf{Validity of the expansion and candidate materials}:\;
The effective Hamiltonian approach is only valid in a time scale shorter than that of heating. 
In Mott insulators, creation of doublon-hole pairs will make the system conducting and destroy the spin picture. 
There are several sweet spots suitable for Floquet engineering in ultrafast spintronics. 
(i) High frequency regime: When $\Omega$ exceeds both $U$ and $t$, several 
doublon-hole pairs must be created simultaneously, which is a slower process. 
Heating becomes exponentially slow in this case~\cite{Mori17}.
Candidate materials are organic Mott insulators~\cite{Shimizu03} since their energy scale is 
one order smaller compared to cuprates. 
(ii) Sub-gap regime $\Omega<\Delta_{\rm Mott}$:
This is an attractive regime since many Mott insulators have 
gaps around and above 1 eV~\cite{ImadaRMP98}, and 
one can use mid-infrared lasers to access the spins.

When the field becomes stronger, charge excitation becomes nonnegligible 
due to higher order processes such as multi-photon absorption, tunneling, 
and even electron avalanche. This is critical for spintronics application, 
but opens a new possibility for a
``photo-induced phase transition", which we explain in the next section.

%%%%%%%%%%%%%%%%%%%%%%%%%%%%%%%%%%%%%%%
\section{Correlated electrons driven by electric fields}
\label{sec:correlated}
%%%%%%%%%%%%%%%%%%%%%%%%%%%%%%%%%%%%%%%

%%%%%%%%%%%%%%%%%%%%%%%%%%%%%%%%%%%%%%%%%%%%%
\begin{figure}[htb]
\includegraphics[width=13cm]{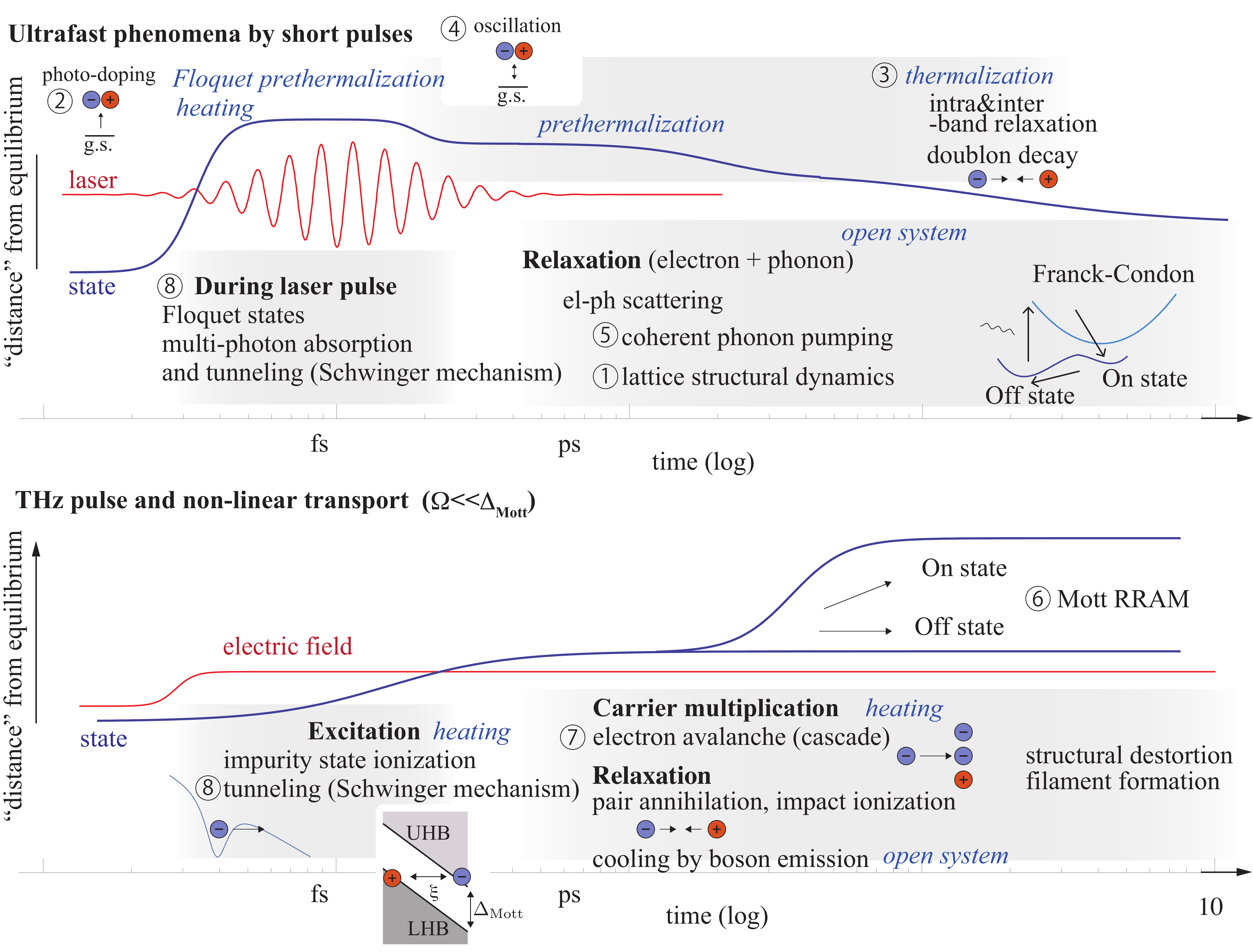}
\caption{Processes taking place in pump-probe experiments with ultra-short pulses 
and with longer pulses or non-linear DC devices (lower panel). 
The blue italic keywords roughly correspond to the many-body processes explained in Section \ref{sec:statistic}.
 }
\label{fig:timescale}
\end{figure}
%%%%%%%%%%%%%%%%%%%%%%%%%%%%%%%%%%%%%%%%%%%%%

Ultrafast phenomena in strongly correlated electron systems driven by intense laser pulses
have been studied during the last decades starting from a pioneering work in organic molecular compounds \cite{Koshihara90,Koshihara92}
and vanadium oxides \cite{Cavalleri01} that are associated with \textcircled{\scriptsize 1}~structural lattice dynamics. 
In many cases, the structural change can be explained through the Franck-Condon picture~\cite{Nasubook};
When electrons are excited, the lattice is subject to the Hellmann\UTF{2013}Feynman force different from equilibrium and 
is moved from the original structure (OFF) to a metastable excited state (ON).
This switching can be used to realize an ``optical memory". 
In cuprates the concept of \textcircled{\scriptsize 2}~``photo-doping", i.e., creating doublon (electron)-hole pairs 
as transient carriers by laser, was introduced~\cite{Yu91}. 
Since then, the time resolution, intensity and the flexibility to adjust the 
photon energy $\Omega$ of the pump laser have improved drastically [reviewed in \cite{Nasubook,Iwai06,Koshihara06,Yonemitsu08,okaLMP,Basov11,Aoki14,Giannetti16,Mankowsky16,Basov17,Cavalleri18review}]. 

Let us explain the sequential processes that take place during and after the pulse laser 
with the corresponding numbers shown in \Fig{\ref{fig:timescale}} with the general 
Floquet many-body physics explained in Section \ref{sec:statistic} in mind.
First, we note that the physics strongly depends on the pulse duration, ranging from 
few femto seconds in the near infrared regime ($\Omega\sim 1.5\mbox{eV}$)
to pico seconds ($10^3$ fs) for THz lasers ($1\mbox{THz}=4.1\mbox{meV}$). 
The typical timescale of electron dynamics is femto seconds, and thus for electrons irradiated by 
THz lasers the duration of the pulse is long enough that 
we expect to see the DC-field physics realized in transistor-like devices with applied bias voltage \cite{Sawa08,Cario10,Pan14}. 

Using ultrashort pulses (\Fig{\ref{fig:timescale}} upper),  it is now possible to observe quantum coherent dynamics in 
strongly correlated electron systems such as \textcircled{\scriptsize 3}~ultrafast switching and relaxation towards a Mott insulator~\cite{Iwai2003} 
or toward a metal~\cite{Perfetti07} and 
it is even possible to see the  \textcircled{\scriptsize 4}~interference oscillation between the Mott insulating groundstate and the 
excited state with a doublon-hole pair~\cite{Wall2011}. 
Interestingly, almost at the same time, 
relaxation dynamics and doublon decay in a strongly interacting fermionic system
was observed in cold atoms in optical traps, 
and this was realized by dynamically changing the trap~\cite{Strohmaier10}. 
The dynamical control of lattice structure is also used in solid states, 
and this is done by exciting \textcircled{\scriptsize 5}~coherent phonon oscillations to optimize the lattice structure 
with an aim to drive the electronics into interesting nonequilibrium phases such as a superconductor~\cite{Fausit11,Hu14,Kaiser14,Cavalleri18review}. 

In the DC limit (\Fig{\ref{fig:timescale}} lower) accessible by THz pump or non-linear transport devices, 
the electrons will be excited continuously and at the same time experience various relaxation processes,
and their balance may realize interesting nonequilibrium steady states. 
The main relaxation mechanism is through emission of bosons, {\em e.g.} phonon,  photon and spin fluctuation.  
In DC devices made from strongly correlated materials, one goal is to realize a
\textcircled{\scriptsize 6}~Mott RRAM (resistivity random access memory);
A device with an $IV$-characteristics showing switching behaviors~\cite{Tokura88,Asamitsu97,Taguchi00,Sawa08,Cario10,Pan14}.
This can be considered as a DC limit of the optical switch. 
The electron avalanche~\textcircled{\scriptsize 7}~is an important excitation mechanism
that makes the carrier density to exponentially increase leading to switching to a conducting state~\cite{Guiot12} 
(also demonstrated in a THz laser-excited semiconductor~\cite{Hirori2011}).

It is still not easy to experimentally study the \textcircled{\scriptsize 8}~electron 
dynamics during the pulse duration in strongly correlated 
materials since the typical pulse duration is few to tens of femtoseconds.
On the other hand,
theoretical studies using the driven Hubbard model (\Eq{\ref{eq:tdHubbard}}) have been done extensively
\cite{Oka03,Okadmrg05,Oka08Luttinger,Takahashi08,Okabethe,Tsujibandflip,Oka12}. 

\noindent
\textbf{(a) Laser driven conducting state at resonance ($\Omega\sim\Delta_{\rm Mott}$)}:\;
In the presence of an AC electric field resonant with the Mott gap, 
the system evolves to a photo-induced metallic state within a short time scale. 
The doublon and hole pairs are resonantly created by the field, 
and at the same time, will be destroyed through stimulated emission,
and the system reaches a transient metastable state~\cite{Oka08Luttinger}. 
In 1D, this state shares common feature as the equilibrium Tomonaga-Luttinger liquid
such as spin charge separation.
Such a metallic state can be captured by combining the Floquet method with the fermion-boson correspondence~\cite{Oka08Luttinger} 
or using the Floquet-Schrieffer-Wolff transformation~\cite{Bukov16}.

\noindent
\textbf{(b) Dielectric breakdown and Keldysh crossover ($\Omega<\Delta_{\rm Mott}$)}  \cite{Oka03,Okadmrg05,Oka12}:\;
Starting from the Mott insulating groundstate, the application of sub-gap AC electric fields will 
trigger dielectric breakdown through quantum tunneling or multi-photon absorption. 
The Loschmidt echo captures how a state $|\psi(t)\ket=e^{-iHt}|\psi(0)\ket$ 
is excited from the initial state, and is defined by their overlap $\bra\psi(0)|e^{-iHt}|\psi(0)\ket$. 
If the overlap decays as $\bra\psi(0)|e^{-iHt}|\psi(0)\ket\sim e^{-i\mathcal{L}'t-\mathcal{L}''t}$ ($\mathcal{L}',\mathcal{L}''$: real),
then $\Gamma=2\mathcal{L}''$ characterizes the decay of the initial state, while $\mathcal{L}'$ is the Aharonov-Anandan phase~\cite{AharonovAnandan}. 
In quantum electrodynamics (QED) in an external field background, 
the Loschmidt echo was studied, and defines the 
Heisenberg-Euler-Schwinger effective 
Lagrangian~\cite{HeisenbergEuler,Schwinger51,Dunne04review,Tanjireview}. 
This was adapted to (interacting) lattice models using the groundstate-to-groundstate 
amplitude~\cite{Okadmrg05,okaLMP}
\begin{eqnarray}
\mathcal{L}(F)=-\lim_{\tau\to\infty}\frac{i}{\tau V}\ln\bra 0|\hat{T}e^{-i\int_0^\tau F(s)\hat{X}(s)ds}|0\ket,
\label{ggamplitude}
\end{eqnarray}
where $F=A\Omega$ is the electric field, $V$ the volume and $\hat{X}=\sum_i in_i$ 
is the position operator in the 
interaction representation $\hat{X}(t)=e^{itH_0}\hat{X}e^{-itH_0}$, where the 
original lattice Hamiltonian $H_0$, {\em e.g.} Hubbard Hamiltonian, is used. 
The real part of the effective Lagrangian is related to the field induced electron polarization 
$
P(F)=\frac{\partial}{\partial F}\mbox{Re}\mathcal{L}(F)$
and reduces to the Berry phase formula~\cite{RevModPhys.66.899,PhysRevB.47.1651} in the weak DC field limit. 
The imaginary part of the effective Lagrangian gives
the decay rate $\Gamma(F)/V=2\mbox{Im}\;\mathcal{L}(F)$
of the insulating ground state. 
The leading contribution to $\Gamma(F)/V$ comes from the doublon-hole creation, 
a process known in QED as the Schwinger effect \cite{Dunne04review,Tanjireview}. 
For the 1D Hubbard model, one can use Bethe ansatz of the non-Hermitian Hubbard model
combined with the imaginary time method
to evaluate the creation probability of doublon-hole pairs \cite{Okabethe,Oka12}
\begin{eqnarray}
\mathcal{P}_{\rm dh}\sim \left\{
	\begin{array}{cc}
\left(\frac{F_0\xi}{h\Omega}\right)^{2\frac{\Delta_{\rm Mott}}{\Omega}}
&\gamma\gg 1 \mbox{(multiphoton) },\\
\exp\left(
-\frac{\pi}{2}\frac{\Delta_{\rm Mott}}{\xi F_0}
(1-\frac{\pi}{16}\gamma^2+\ldots)
\right)&\gamma\ll 1 \mbox{(tunneling) }
\end{array}
\right.
\label{crossover}
\end{eqnarray}
for AC fields $F(t)=F_0\cos\Omega t$. 
The correlation length $\xi$ characterizes the size of the doublon-hole pairs existing in the 
groundstate as a quantum fluctuation~\cite{PhysRevB.48.1409}, and the Keldysh parameter 
which separates the tunneling dominant to multiphoton dominant 
creation is defined by $\gamma=\frac{\Omega}{F_0\xi}$. 
Reaching and exceeding the Schwinger limit $F_{\rm Sch}=\frac{\Delta_{\rm Mott}}{\xi}$ 
is a challenging problem, and typically this is impossible since 
the electron avalanche occurs below the limit~\cite{Guiot12,Liu12} 
(also known in QED~\cite{Fedotov10}). 
Recently, the Schwinger limit was reached and exceeded in correlated insulators,
and the tunneling breakdown (\Eq\ref{crossover} lower) was experimentally verified~\cite{Mayer15,Yamakawa17}. 
These outstanding result became possible
by using a laser pulse that is short enough to suppress the heating and avalanche cascade effect.

%%%%%%%%%%%%%%%%%%%%%%%%%%%%%%%%%%%%%%%
\section{Outlook}
\label{sec:outlook}
%%%%%%%%%%%%%%%%%%%%%%%%%%%%%%%%%%%%%%%
One of the most fascinating aspect of Floquet engineering is that it 
is becoming a common language for researchers with diverse backgrounds 
such as strongly correlated electron systems, artificial matter and nonequilibrium statistical mechanics. 
High energy physics and quantum field theory~\cite{HeisenbergEuler,Schwinger51,Dunne04review,Tanjireview} 
is also a great source of ideas and motivations for condensed matter physicists and the interaction 
is expected to fertilize the field. 
As a closing, let us mention some challenges which may be interesting for future developments.  

\noindent
\textbf{Tailored fields from metamaterial plasmonics}:\;
In order to make the Floquet engineered devices to fit inside our pockets, we need 
to replace the laser with a more efficient field generator, and
the usage of metamaterials and near field optics will be an important step in this direction. 
Although it must be triggered by a pulse laser, 
metamaterials have demonstrated that an electric fields exceeding 1MV/cm~\cite{Liu12}
or a magnetic field of 1 Tesla~\cite{Mukai16}, oscillating in the THz regime can be generated. 
Interestingly, the fields are ``tailored" by the structure and is different from 
the traveling EM-field in the vacuum. It would be interesting to look for 
Floquet states that utilizes this freedom.

\noindent
\textbf{Phase transition in a nonequilibrium state}:\;
Strongly motivated by recent experiments on light-induced superconductivity~\cite{Fausit11,Hu14,Kaiser14,Cavalleri18review}
several theoretical researches are done in driven correlated open systems~\cite{Knap16,Murakami172,Babadi17}. 
The DC counterpart, {\em i.e.}
phase transitions that occurs in non-linear devices~\cite{Sawa08,Cario10,Pan14,Rozenberg04,Lee14}, also 
show interesting properties~\cite{Chanchal17},  and provides challenges to theorists.

\noindent
\textbf{New device from new principle}:\;
It is interesting to look for functions of Floquet states that 
 static systems cannot have. 
Devices that have an output signal with non-trivial frequencies due to the 
dynamics of the Floquet state have been proposed 
using correlated electrons~ \cite{Silva18,MurakamiHHG}, 
two dimensional electron gas~\cite{Oka16}, and even single spins~\cite{Martin17}.

%Disclosure
\section{DISCLOSURE STATEMENT}
The authors are not aware of any affiliations, memberships, funding, or financial holdings that
might be perceived as affecting the objectivity of this review.

%%%%%%%%%%%%%%%%%%%%%%%%%%%%%%%%%%%%%%%
%\subsection{New states, new functions}
%\label{sec:newfunction}
%%%%%%%%%%%%%%%%%%%%%%%%%%%%%%%%%%%%%%%

%%%%%%%%%%%%%%%%%%%%%%%%%%%%%%%%%%%%%%%
\section{ACKNOWLEDGMENTS}
\label{sec:ACKNOWLEDGMENTSk}
%%%%%%%%%%%%%%%%%%%%%%%%%%%%%%%%%%%%%%%
We gratefully acknowledge Naoto Tsuji, Takuya Kitagawa, Hideo Aoki,  Shintaro Takayoshi,
Masahiro Sato, Aditi Mitra, Hossein Dehghani, 
Takahiro Mikami, Eugene Demler, Liang Fu, Andr\'e Eckardt and Achilleas Lazarides
for useful discussions, and financial support from the Max Planck Society.

%\bibliographystyle{apsrev.bst}
%\bibliographystyle{apsrmp.bst}
%\bibliographystyle{unsrt.bst}
%\bibliography{c:/Physics/ref}
%\printindex

\bibliographystyle{ar-style4.bst}
\bibliography{reference}

\end{document}